\documentclass[11pt,a4paper]{article}
\pdfoutput=1
\usepackage[english]{babel}
\usepackage[latin1]{inputenc}
\usepackage{amsfonts,amsbsy,bm,euscript,mathrsfs}
\usepackage{amssymb,stmaryrd,faktor,slashed}%
\usepackage{color}
\usepackage[tbtags]{amsmath}
\usepackage[bookmarks=true,colorlinks=true,linkcolor=black,
citecolor=black,urlcolor=black,bookmarksnumbered]{hyperref}

\usepackage{amsmath,amsfonts,mathrsfs,graphicx,bbm}
\usepackage{fixltx2e}
\usepackage{cancel}

\numberwithin{equation}{section}

\makeatletter
\renewcommand\section{\@startsection {section}{1}{\z@}
{-3.5ex \@plus -1ex \@minus -.2ex}
{2.3ex \@plus.2ex}
{\normalfont\Large\bfseries}}
\renewcommand\subsection{\@startsection{subsection}{2}{\z@}
{-3.25ex\@plus -1ex \@minus -.2ex}
{1.5ex \@plus.2ex}
{\normalfont\large\bfseries}}
\makeatother

\newcommand{\alg}[1]{{\mathfrak{#1}}}
\newcommand{\su}{\alg{su}}

\newcommand{\psu}{\alg{psu}}

\newcommand{\op}{\mathcal{O}}

\newcommand{\gen}{\mathbf}

\def\vp{\varphi}
\def\vk{\varkappa}

\def\T{\Theta}
\def\z{\zeta}

\def\r {\rho}
\def\G{\Gamma}

\def\pa {\partial}
\def\CF{f}

\newcommand{\STr}{\operatorname{Str}}

\def\mf{\mathfrak f}

\def\bg{\boldsymbol{\gamma}}

\newcommand{\arxivlink}[1]{\href{http://arxiv.org/abs/#1}{arXiv:#1}}
\newcommand \foot [1] {\footnote{#1\vspace{2pt}}}
\newcommand \rf [1] {(\ref{#1})}

\def \be {\begin{eqnarray}}
\def \ee {\end{eqnarray}}

\def \STr {{\rm STr}}

\def \bi{\bibitem}

\def \ha {{1 \over 2}}
\def \td {\tilde}
\def \ci{\cite}

\def \z {\zeta}

\def \a {\alpha}
\def \b {\beta}

\def \del {\partial}
\def \a {\alpha}

\def \g {\gamma}
\def \s {\sigma}
\def \z {\zeta}

\def \ov {\over}

\def \b {\beta}
\def \l {\lambda}

\def \ci {\cite}

\def \P {\Phi}
\def \l {\lambda}

\def \M {{\mathcal M}}

\def \td {\tilde}

\def \m {\mu}

\def \bi{\bibitem}
\def \la {\label}
\def \l {\lambda}

\def \adss {$AdS_5 \times S^5~$ }
\def \ov {\over}

\def \F {{\cal F}}

\def \ha {{1\ov 2}}
\def \r {\rho}
\def \no {\nonumber}

\def \del {\partial}

\def \bi {\bibitem}
\def \la {\label}
\def \l {\lambda}

\def \adss {$AdS_5 \times S^5$\ }

\def \r {\rho}

\def \ov {\over}

\def \varpi {{\rm w}}

\def \pa{\partial}

\def \ep {\epsilon}

\def \s {\sigma}

\def \n {\nu}

\def \vp {\varphi}

\def \ha {{{\textstyle{1 \ov2}}}}
\def \fo {{\textstyle{1 \ov4}}}

\def \eqref {\rf}
\def \ads {$AdS_3 \times S^3$\ }
\def \adss {$AdS_5\times S^5$\ }

\def \iffa {\iffalse}

\def \Str {{\rm Str}}

\def \ads {$AdS_5 \times S^5$\ }

\def \adst {$AdS_3 \times S^3$\ }

\def \emo {$\eta$-model\ }
\def \lmo {$\l$-model\ }

\def \vk {\varkappa}
\def \hh {h}

\def \ddss {{\oplus}}

\def \SS {{\mathcal S}}
\def \G {\Gamma} 
\def\T{\Theta}
\def\z{\zeta}

\def\r {\rho}
\def\G{\Gamma}

\def \vka  {\vk}  \def \vkappa  {\vk}
\def \cla {{\rm c}}
\def \G {\Gamma} 

\begin{document}

\setcounter{equation}{0}
\setcounter{footnote}{0}
\setcounter{section}{0}

\vspace{-3cm}
\thispagestyle{empty}
\vspace{-1cm}

\rightline{ Imperial-TP-RB-2016-01}

\rightline{ \today}

\begin{center}
\vspace{1.5cm}

{\Large\bf  Supergravity background    \\  \vspace{0.3cm} 
of $\l$-deformed  model     
   for   $AdS_2 \times S^2$ supercoset  
}

\vspace{1.5cm}

{R. Borsato\footnote{r.borsato@imperial.ac.uk}, 
A.A. Tseytlin\footnote{Also at Lebedev Institute, Moscow. tseytlin@imperial.ac.uk }
and L. Wulff\footnote{l.wulff@imperial.ac.uk   }
}
\vskip 0.4cm

{\em  The Blackett Laboratory, Imperial College, London SW7 2AZ, U.K.}

\vspace{.2cm}
\end{center}

\def \hF {{\widehat {F}}}

\begin{abstract}
Starting   with the $\hF/G$   supercoset model corresponding to the $AdS_n \times S^n$  superstring 
one  can define the  $\l$-model   of arXiv:1409.1538  either as a deformation of the
$\hF/\hF$  gauged WZW  model or as an  integrable  one-parameter   generalization 
of the non-abelian T-dual  of the $AdS_n \times S^n$  superstring   sigma model 
with respect to the whole supergroup $\hF$. 
Here we consider  the case of $n=2$  and find the explicit form of the 4d  target space background 
 for  the  
  $\l$-model for the     $PSU(1,1|2)/SO(1,1) \times SO(2)$  supercoset. 
    We show that  this background represents  a solution  of  type IIB   10d supergravity compactified
     on a 6-torus with only metric, dilaton $\P$ and  the 
 RR 5-form (represented by a  2-form $F$ in 4d)
  being non-trivial. 
 This   implies that  the $\l$-model   is     Weyl invariant  at the quantum level 
 and thus defines a consistent superstring   sigma model. 
 The supergravity  solution  we find  is different from the one  in arXiv:1410.1886  which  
should   correspond to a  version of the $\l$-model   where only the bosonic subgroup of $\hF$
 is  gauged. 
 Still, the two  solutions  have   equivalent  scaling limit 
 of arXiv:1504.07213  leading to the  isometric  background  for the metric and $e^\P F$ 
 which   is  related 
  to the $\eta$-deformed $AdS_2 \times S^2$   sigma  model of arXiv:1309.5850.
 Similar    results   are expected in the 
 $AdS_3 \times S^3$  and $AdS_5 \times S^5$   cases. 
\end{abstract}


\def \k {\varkappa}

\def \emo {$\eta$-model\,}
\def \lamo {$\lambda$-model\,}
\def \h {\kappa}
\def \ed {\end{document}}
\def \P {\Phi} 
\def \hi {\h^{-1}}
\def \F {{\cal F}}
\def \we {\wedge}

\def \Ti {T^{-1}}

\newpage
\setcounter{equation}{0}
\setcounter{footnote}{0}
\setcounter{section}{0}

\def \ads {$AdS_n \times S^n$\, }

\def \adst {$AdS_2 \times S^2$\, }
\def \tex {\textstyle}
\tableofcontents

\section{Introduction}\label{secint}

There   are two special integrable models
that are closely associated with the superstring sigma model on $AdS_n \times
S^n$.  One is   the \emo 
 \ci{dmv}  -- a particular integrable deformation of the \ads
supercoset model   generalising the bosonic Yang-Baxter sigma
model of \ci{Klimcik:2002zj}. 
The other one  is the \lmo 
\ci{hms1,hms2}  generalising the bosonic model of \ci{Sfetsos:2013wia} (see also \ci{tse}).
The \lmo 
is based on the $\hF/\hF$ gauged WZW model 
 closely related  to the \ads supercoset  and  may
be interpreted as an integrable deformation of the non-abelian T-dual of the
\ads supercoset action.

While  for the \emo the corresponding target space background was  found    in 
\ci{abf1,hrt,abf2} (but turns out not to be a supergravity solution \ci{us3}), in the
  case  of the \lmo   the GS   sigma model   action  was  so far   not 
determined  directly 
 apart from the metric \ci{ST,ST2}   and the dilaton \ci{hms2,us1}. 
Our aim below   will be to find  the full \lmo   background (metric, dilaton {\it and} the RR  field strength) 
 from the \lmo  action  and also as a    solution of the type II supergravity   equations. We shall   consider  
  the  simplest  example of the \adst model.  The resulting background  differs from 
 the supergravity solution  based on the  metric and  dilaton of the bosonic  model  that was found   in \ci{ST}. 

Let us start  with a brief review of the \lmo  \ci{hms2} (see  also  \ci{us1}). 
The \lmo 
 may be interpreted as a  unique integrable deformation of the    first-order  action that    interpolates 
 between the  supercoset \ads   superstring model  and its  non-abelian T-dual   with respect to the full supergroup symmetry. 
In general, one may consider a model  based on the supercoset
$
\frac{\widehat{F}}{G_1 \times G_2} \supset \frac{F_1 }{G_1}\times \frac{F_2}{G_2}\,
$
where $\widehat{F}$ is a supergroup (e.g. $PSU(2,2|4)$ in the $AdS_5 \times S^5$ case or $PSU(1,1|2)$ in the \adst case)
and $F_i$ and $G_i$ are bosonic subgroups.
The \lmo  is defined by  the action
\be
&&\hat I_{k,\l}(f,A)=\frac{k}{{4}\pi} \Big(
\int d^2 x \; \operatorname{STr}\big[\tex  \frac12 f^{-1}\partial_+ f f^{-1}\partial_- f + A_+ \partial_- f f^{-1}
\no \\&&\qquad \qquad 
- A_- f^{-1}\partial_+ f - f^{-1} A_+ f A_- + A_+ A_- \big] \no \\ &&
-{\tex  \frac{1}{{3}}} \int d^3x \; \epsilon^{abc}\operatorname{STr}
\big[f^{-1}\partial_a f f^{-1}\partial_b f f^{-1}\partial_c f\big]
\label{1}
+ (\lambda^{-2} - 1) \int d^2 x \; \operatorname{STr}\big[A_+ P_\lambda A_- \big]\Big)\ ,
\ee
where $f \in \widehat{F}$, \ $A_\pm \in \hat{\mathfrak{f}}$=alg$(\widehat{F})$  and $P_\lambda $ is a combination of $\mathbb Z_4$   projectors\foot{Equivalently, 
$(\lambda^{-2} - 1) A_+ P_\lambda A_- =   A_+ ( \Omega - 1) A_-$,  where 
$\Omega=  P^{(0)}   + \l^{-2} P^{(2)}  + \l^{-1} P^{(1)}  + \l P^{(3)}$.} 
\begin{equation}\la{pii}
P_\lambda = P^{(2)}  + \frac1{\lambda^{-1}+1} (P^{(1)}  -\lambda P^{(3)} ) \ .
\end{equation}
All but the last term in  \eqref{1} correspond to the $ \widehat{F}/\widehat{F}$ gauged WZW model with integer  coupling
(level) $k$ and $\l$ is a ``deformation" parameter.
This action has no  global symmetry but there is a 
 local fermionic symmetry and 
a $G_1 \times G_2$ gauge
symmetry which   will be  fixed   by a condition on $f$  after integrating out  the gauge fields.

The direct  limit $\l \to 1$ for fixed $k$   leaves one with $ \widehat{F}/\widehat{F}$ gauged WZW model. 
One can also consider   another  special 
limit of $\l\to 1$ combined with  sending $k\to \infty$   and   scaling  the supergroup  field  $f\to 1$ as in  \ci{Sfetsos:2013wia}
\begin{equation}\label{natdlim}\tex 
f= \exp(-\frac{{4}\pi}{k}\, v )= 1 - \frac{{4}\pi}{k}v + \mathcal{O}(k^{-2}) \ , \qquad \lambda =
1-\frac{\pi }{k} h + \mathcal{O}(k^{-2}) \ , \qquad k \to \infty \ ,
\end{equation}
where   $v\in \hat{\mathfrak{f}}$ and  $\hh$ are kept fixed. This
leads to the following action
\begin{equation} \la{9}
\hat I_{k\to \infty,\l\to 1}(f\to 1,A)= \int d^2x \; \operatorname{STr} \big[ \, v \, (\partial_- A_+ - \partial_+ A_- + [A_-,A_+])\big]
+ \ha {\hh} \int d^2 x \; \operatorname{STr} \big( A_+ P A_- \big)\ ,
\end{equation}
where $P=P_\l \big|_{\l=1}= P^{(2)}  + \ha ( P^{(1)}  - P^{(3)} )$    is the projector  appearing in the standard  \ads   superstring 
Lagrangian $L= \operatorname{STr} \big[ J^{(2)}_+ J^{(2)}_-   + \ha ( J^{(3)}_+ J^{(1)}_-  - J^{(1)}_+  J^{(3)}_-)\big]$ \ \ci{mts,bersh}.
Eq. \rf{9} may  be interpreted as a first-order action ``interpolating" between the
 \ads supercoset action (if one first integrates out $v$ getting  $A_\pm = g^{-1} \del_\pm g$) and its 
 non-abelian T-dual   with respect to the full supergroup  (if one first integrates out $A_\pm$).\foot{Here  the   non-abelian duality  contains both the standard   bosonic  and   also   the fermionic  transformations    like in the abelian fermionic T-duality in  \ci{berk}.}
Thus the \lmo  \rf{1}  may be interpreted as a  deformation of the first-order interpolating action \rf{9}.
If one first integrates out $A_\pm$ in \rf{1} and gauge-fixes the supergroup field  $f$
 the resulting sigma model may be viewed as a deformation of the non-abelian T-dual of the original  \ads supercoset model.
\foot{Another special   limit of \rf{1}  is $\l\to 0$     in which $A_+ ( \Omega - 1) A_- \to 
A_+ ( \l^{-2} P_2 + \l^{-1} P_1 - P_3 ) A_-$  implying that we should set  
$A_\pm^{(2)}=0, \ A_-^{(1)}=0, \ A_+^{(3)}=0$  so that  the remaining gauge fields  are from  the bosonic  subalgebra 
and  ``half" of the fermionic  directions (reflecting  the presence of $\k$-symmetry in the  model resulting upon integrating out the remaining $A_\pm^{(0)},  \ A_+^{(1)}, \ A_-^{(3)}$  fields). 
Integrating out  gauge fields   and fixing gauge on $f$  will still lead to a non-trivial  background discussed below.}

\iffa While there is a close on-shell connection between the models \eqref{lagdmv}
and \rf{1} at the level of classical Hamiltonian (Poisson-bracket)
structures \ci{dmv,hms1,hms2}, establishing their 
correspondence at the level of
the actions (and thus eventually at the quantum level) remains an open problem
that we will attempt to address below.\foot{Note that integrability, together
with expected quantum UV finiteness, suggest that classical relations may in
some way extend to the quantum level.}\fi 


Next, let  us recall the relations  between parameters of the \emo and \lamo  \cite{dmv,hms2}. 
In terms of the Poisson algebra deformation parameter $\ep$  the parameter $\eta$ of \cite{dmv} (or $\varkappa$ introduced
in \cite{abf1}) is 
\begin{equation}\label{releps1}
\epsilon^2 = \frac{4 \eta^2}{(1+\eta^2)^2} = \frac{\varkappa^2}{1+\varkappa^2} \ ,\qquad
\qquad \epsilon^2 \in [0,1] \ , \quad \eta^2 \in [0,1] \ , \quad \varkappa^2 \in [0,\infty] \ ,
\end{equation}
The  parameter $\l$ in the action \rf{1} of \cite{hms2} is related
to $\epsilon^2$ by
\begin{equation}\begin{split}\label{releps2}
\epsilon^2 = -\frac{(1-\lambda^2)^2}{4\lambda^2} = -\frac{1}{4b^2(1+b^2)}\ ,\qquad
\qquad & \epsilon^2 \in [-\infty,0] \ , \quad \lambda^2 \in [0,1] \ , \quad b^2 \in [0,\infty] \ .
\end{split}\end{equation}
Here $\vk={ 2 \eta \ov 1-\eta^2}$   and 
$b^2 = \frac{\lambda^2}{1-\lambda^2} $
are  natural deformation parameters in the bosonic parts  of the two models. 
Comparing \rf{releps1} and \rf{releps2} 
the parameters of the two deformed models
may be related by an analytic continuation \ci{us1} 
\be \la{218}
&&\eta = i { 1 - \l \ov 1 + \l } \ , \ \ \qquad\qquad \l = {i - \eta \ov i + \eta} \ , \la{1s} \\
&& \ \vk =  i \h \ , \la{219}
\qquad\qquad   \h\equiv 
{ 1 \ov 1 + 2 b^2 } = \frac{1-\lambda ^2}{1+\lambda ^2} \  , \ \ \ \ \ \ \ \ 
\l = \sqrt{ 1 -\h \ov 1+ \h }  \ ,
\ee
where $\h \in [0,1]$ (for $\lambda^2 \in [0,1]$) 
 is the   parameter   that  we will often use   below  instead of $\l$. 
Also,  the overall couplings of the two models are related by ($h$ is the  tension  of the \emo)\foot{The relation between the
quantum deformation parameters $q$ for the two models (cf. \ci{dmv,abf1,hms1,hms2,us1}) is 
$ q= e^{ - {\k \ov h}} \leftrightarrow q= e^{ - {i\pi \ov k}} \ ,  $
with the real $q$ corresponding to the \emo and the root of unity $q$ to the $\l$-model.
}
\be \la{18}
\ \ \ { k \ov \pi} = i {h\ov \vk} \ , \ \ \ \qquad {\rm i.e. } \qquad \ \ \ h= { k \ov \pi  }\,  \h \ .
\ee
As was  found  in \ci{us1},   the  relations  \rf{1s},\rf{18}     allow one to   obtain 
the metric of the    \emo  as a special limit  
of the \lmo  metric   \rf{1}.
More precisely, this singular   limit (that generates isometries  corresponding  to  the bosonic Cartan  directions)
leads to  an abelian T-dual of the \emo   metric \ci{us1}. 

Starting   with the bosonic   version   of the \lmo in  \rf{1}  corresponding to the $AdS_n \times S^n$ supercoset 
 and integrating  out  the  gauge fields  $A^{(2)}_\pm$  
  one  can find   the corresponding  metric and dilaton $\P_B$ field  \ci{ST,ST2}. 
In   \ci{ST,ST2}  this  background  was embedded  as a solution into     type II  supergravity    by  finding the 
corresponding   RR field strength. 
The    limit  \ci{us1}     of this    supergravity background   was shown   \ci{us2}   to give a   type II supergravity 
solution   which  has  the metric and  RR   field  $\F= e^{\P_B}   F$  which are related by the 
 standard T-duality rule  to the 
metric and $\F$     extracted from the \emo action in \ci{abf1,abf2}. 
However, this  scaling  limit  leaves  a term in the  dilaton  which is linear  in isometric coordinates   thus obstructing 
 the application of the standard T-duality  to  the full background. 
This is an explanation for why the \emo    background does not correspond to a solution of supergravity. Indeed, it was   found to solve only  the  weaker 
one-loop scale invariance conditions  but not the Weyl invariance   conditions (equivalent to the  supergravity equations) 
for the corresponding superstring sigma model \ci{us3}.

Starting   instead   with  the full supercoset   \lmo  in \rf{1}   and integrating out 
 both the bosonic and fermionic   components 
of the gauge fields  $A_\pm$   one  gets the same  effective sigma model   metric as in the  bosonic   model case \ci{ST}
but the expression for the dilaton  turns out to be different from the ``bosonic" one in \ci{ST,ST2} 
 containing an extra ``numerator"  factor  
 from  integration over  the   fermionic  components of the  gauge fields \ci{hms2,us1}. 
 As we shall   show  below on  the  \adst   supercoset  example,   the resulting metric-dilaton background 
also  solves  the type II  supergravity equations when supplemented by a proper  RR field strength  $F$   which is different from the one   in \ci{ST}. Thus   the   same \lmo   metric can   be embedded into type II  supergravity using   at least two different 
 $(\P,F)$   pairs. Similar   non-uniqueness  of the  supergravity solutions   was observed in \ci{lrt}   in the \emo  context. 
 
 
 Furthermore,  we shall show that it is the  combination  $\F=e^\P F$  of this   RR field strength   
   with the \lmo  dilaton  \ci{us1}    that     indeed   directly  corresponds to the 
  sigma model  that originates from the \lmo \rf{1}, i.e. this  background is   the one  that 
  corresponds to the \lmo  of \ci{hms2}.  The fact that this background solves the   supergravity 
  equations    confirms  that the \lmo  is not only scale-invariant \ci{holl}   but (in contrast to the  \emo \ci{us3})   is  also 
   Weyl-invariant  as a quantum  sigma model  on a curved 2d background  and thus  
   it    defines     a  consistent superstring theory. 
 
The scaling limit  \ci{us1} applied to this    new  solution   leads to  an equivalent 
 background  to the one   found  in  \ci{us1}  from  the scaling limit    of the ``bosonic" solution, 
 in agreement with  the expected  relation between  the \lmo and \emo.\foot{The equivalence between the scaling limits 
 of the ``bosonic" background  that should correspond to gauging just bosonic  generators 
 and the full \emo  background   may be understood   by noting that  the scaling limit  ``blows up"   the  bosonic  Cartan directions, 
 and thus    the gauging of the fermionic directions   should not be important.}
 
 \

The  structure  of this paper is  as follows. 
 We shall start in section 2   with  reviewing   the  form of the  4d metric and dilaton  corresponding to the \lmo  for the \adst   supercoset. 
 We shall then  present   the solution of  type IIB   supergravity compactified on a 6-torus that supports  this  metric  and dilaton 
 background by a RR 5-form  background. 

 In section 3 we shall   explain how to extract the metric, dilaton {\it and} this  RR  background    
 \rf{1}  by  integrating out  the gauge fields $A_\pm$  in the \lmo action  \rf{1},  writing  the resulting quadratic fermionic action in the  GS    superstring   sigma model form   and using the $\k$-symmetry   invariance \ci{hms2}    of the resulting action. We shall use a short-cut method 
 based on studying the structure of the $\k$-symmetry variation of the world-sheet metric. 

Section 4    will contain some concluding remarks. 
In Appendix \ref{apa}    we will   check that   the   \lmo background  is a solution of type IIB supergravity 
directly in 10 dimensions  starting with  the type II   equations of motion written in  bispinor  notation  for the RR   field strengths. 
Appendix \ref{apb}  will 
present  the realisation of $\su(1,1|2)$  used in section 3.
Appendices \ref{apc}   and \ref{apb}  will   contain some technical details on $\k$-symmetry variations 
  and   representation of Dirac matrices.

\section{ \lmo  background  for the   $AdS_2 \times S^2$  supercoset } 
\label{sec2}

\subsection{Metric }\label{sec21}
To find the target space  metric it is sufficient to set fermions to zero, i.e. consider just the bosonic  version of the \lmo  \ci{ST}.  
In the case of $AdS_2 \times S^2$
the relevant bosonic coset space is
$\frac{SO(2,1)}{SO(1,1)} \times \frac{SO(3)}{SO(2)} $.
Starting with the \lmo action \rf{1},
integrating out the gauge field and imposing a  gauge-fixing  condition on the $SO(2,1) \times SO(3)$ field $f$ by choosing it 
as  \ci{us1} 
\begin{equation}\label{patch1}
f= \big[\exp(i t \sigma_3) \exp(\xi \sigma_1)\big] \ddss \big[ \exp(i \varphi \sigma_3) \exp(i\zeta\sigma_1)\big] \ ,
\end{equation}
we find the following metric\foot{We shall use 
similar  notation  for the bosonic part of the
action  as in \ci{us1}: \ \ 
$I=  \ha \int d^2 x \ g_{mn} (X) \del_+ X^m \del_- X^n $ with
$ds^2 = g_{mn}(X) d X^m d X^n$, i.e.   
the tension $T= { k\ov \pi} $  in the \lmo or  $T= {h}  $ in the \emo
will be included in the metric.}
\begin{equation}\label{lag1}\begin{split}
T^{-1}  ds^2 =\h \big[ & -dt^2 + \cot^2 t \, d\xi^2
- ( \h^{-2} -1)  (\cosh \xi \, dt - \cot t \sinh \xi \, d\xi)^2
\\
& + d \varphi^2 + \cot^2 \varphi \, d\zeta^2
+ ( \h^{-2} -1)  (\cos \zeta \, d\varphi + \cot \varphi \sin \zeta \, d\zeta)^2 \big] \ .
\end{split}\end{equation}
In  \cite{ST}    a different 
coordinate patch  was used   where the metric   is related to the one in \rf{lag1} 
by the   analytic continuation $(t, \xi)\to (\td t, \td \xi)$  with  $t =-i \tilde \xi , \  \xi = \tilde  t $. 
Explicitly,  choosing 
\be\label{patch2}
 f= \big[ \exp(\tilde \xi \sigma_2) \exp(\tilde t \sigma_1)\big]\ddss\big[ \exp(i \varphi \sigma_3) \exp(i\zeta\sigma_1)\big] \ ,
\ee
leads   to
\begin{equation}\begin{split}\label{lag1a}
\Ti \widetilde {ds}{}^2 = \h \big[ &d\tilde\xi^2 - \coth^2 \tilde \xi \, d\tilde t^2
+  ( \h^{-2} -1)   (\cosh \tilde t \, d\tilde \xi + \coth \tilde \xi \sinh \tilde t \, d\tilde t)^2
\\
& + d \varphi^2 + \cot^2 \varphi \, d\zeta^2
+ ( \h^{-2} -1)  (\cos \zeta \, d\varphi + \cot \varphi \sin \zeta \, d\zeta)^2 \big] \ .
\end{split}\end{equation}
The  metric   \rf{lag1a}  (times 6-torus) can be embedded \ci{ST}  into type  II   10d supergravity
  if supplemented   with a  real dilaton and RR 5-form flux $F$ while a  similar embedding 
  of \rf{lag1} requires  an  imaginary RR flux. 
 
  The  ``real"   patch  choice   of   \rf{patch2}   is more natural in the context of
  the full  supercoset  \lmo  and as  we shall see  below 
 the corresponding RR background will again be real for \rf{lag1a}  and imaginary for \rf{lag1}. 
 Note that the metric  \rf{lag1}  or \rf{lag1a}  has no isometries. 
The reason   why \rf{lag1}   was preferred   in \ci{us1}   is that 
it admits a special singular   coordinate  redefinition  in which the first 2d part   of the 4d metric  develops a time-like 
(rather than space-like as for \rf{lag1a}) isometry  and thus is  related 
to  the metric corresponding to the
$\eta$-deformed $AdS_2 \times S^2$ model of \cite{Delduc:2013fga,dmv}.

The two metrics  \rf{lag1}  and \rf{lag1a}   look essentially  the same when written in the algebraic coordinates
 $ (x,y; p,q)$
defined  for \rf{lag1} 
by \ci{us1}\foot{Compared to \ci{us1} we rescaled   these  coordinates by factors of $\h$ 
to make  the metric manifestly conformally flat.
Explicitly, 
$
x=  \h^{-1/2} \cos t\,   \cosh \xi, \  y=  \h^{1/2} \cos t\,   \sinh \xi, \ 
p=  \h^{-1/2}  \cos \vp\,   \cos  \z, \  q=  \h^{1/2}  \cos \vp\,   \sin \z.$
For the coordinate patch  used in \ci{ST}   we have 
$
x=  \h^{-1/2} \cosh \td \xi \,  \cosh \td t , \  y=  \h^{1/2} \cosh t\,   \sinh \td  t, $
so that   instead of $  \h x^2 - \hi y^2  \leq 1 $   we have 
$  \h x^2 - \hi y^2  \geq  1$, i.e. $y$  rather than $x$  is      playing the role  of a time-like   direction.
}
\be
&& t = \operatorname{arccos}\sqrt{\h x^2-\hi y^2} \ , \qquad 
 \xi = \operatorname{arccosh}\frac{\h^{1/2}  x }{\sqrt{\h x^2-\hi y^2}} \ , 
 \qquad \h x^2 - \hi  y^2 \leq1\ ,\qquad 
 \la{1z}
\\
&& \varphi = \operatorname{arccos}\sqrt{\h p^2+\hi q^2} \ , \qquad 
 \zeta = \arccos\frac{\h^{1/2} p}{\sqrt{\h p^2+\hi q^2}}\ , 
 \qquad  \h  p^2 + \hi q^2 \leq1 \ .
 \la{311}
\ee
Then   \rf{lag1}    takes a simple diagonal form
\begin{equation}\la{312}
\Ti  ds^2 = \frac1{1-\h x^2+\hi y^2}\big(- dx^2 + dy^2 \big)
+ \frac1{1-\h p^2-\hi q^2}\big(dp^2 + dq^2\big) \ .
\end{equation}
This  metric  has an asymptotically flat  region and    no isometries.
The metric \rf{lag1a}   is    also given   by \rf{312}
 in the coordinate patch   where 
\be   \h x^2 - \hi y^2  \geq  1   \ , \la{yy} \ee
  i.e. when  $y$ is time-like and $x$  is space-like.\foot{The first 2d part of the  metric 
   \rf{lag1a}    written in similar conformal coordinates appeared also in eq. (5.19) in \ci{ST}.
  Note   that  the parameter $\l$  used  in \ci{ST}  is the 
square of $\l$ used in \ci{hms2,us1} and  here.}

In what follows   we shall  formally use  the metric \rf{312}    with 
an understanding that one can always  consider 
the   physical region   for the \lmo  
 \rf{yy}
  where   \rf{312}   will be supported by a real  dilaton and 
RR background.

\subsection{Dilaton}\label{sec212}

Assuming that the \lmo  defined  by \rf{1}   has no ``bare" dilaton term, 
  the dilaton   should be    generated (as in the  standard  T-duality  transformation or gWZW models)   upon 
   integrating out  the  gauge fields $A_\pm$.   In the purely bosonic \lmo 
one then gets for  \rf{patch1}, i.e. for the metric \rf{lag1},\rf{312} ($e^{ \P_0} = T \h\  e^{\bar \P_0}$)
\be\la{d1}  
e^{\P_B} = e^{\bar \P_0}( \sqrt g)^{1/2}=  \frac{e^{\Phi_0}}{\sin t \ \sin \vp}=   { e^{\P_0} \ov \sqrt{ (1-\h x^2+\hi y^2)   (1-\h p^2-\hi q^2)} }\ . \ee
The   dilaton   corresponding to  \rf{patch2}, i.e. for the metric \rf{lag1a} or \rf{312} in the region 
 $1-\h x^2+\hi y^2 < 0$ 
is found by  the obvious  analytic continuation  leading to a factor of $i$ that   can   be  absorbed into a shift of $\P_0$. 
This  dilaton
\be
 \la{d2}
e^{\P_B} = { e^{\P_0} \ov \sqrt{- (1-\h x^2+\hi y^2)   (1-\h p^2-\hi q^2)} }\  \ee
was   assumed as a starting point  for constructing a supergravity
embedding  for  the metric   \rf{lag1a}   in  \cite{ST}.

At the same time,  integrating out the  full superalgebra gauge field in \rf{1} 
 leads \cite{hms2} 
 to an extra fermionic   $A^{(1)}_\pm,A^{(3)}_\pm$   contribution  to  the dilaton 
 which in the present \adst  supercoset  case  is \ci{us1}   (for the 
group field $f$   having only the bosonic part \rf{patch1})  
\be
&&  \la{d3}
e^{\P} = e^{\P'_0}  M'  ( \sqrt g)^{1/2}=  \frac{e^{\Phi'_0} M' }{\sin t\ \sin \vp}\ , \\
&& 
M'=  - (1-\lambda ^2)^2 +   (1+\lambda^4 + 2  \lambda^2 \cosh 2 \xi )\cos^2 t 
+ (1+ \lambda^4 + 2\lambda^2 \cos 2 \zeta ) \cos^2\varphi  \no 
\\ && \hspace{100pt}-4 \lambda  (1+\lambda ^2) \cos t \cos\vp  \cosh \xi  \cos \zeta\ . \la{d39}
\ee
In terms of the algebraic coordinates and $\h= { 1 -\l^2 \ov 1 + \l^2}$ in \rf{219} the  fermionic contribution   $M'$ is 
\be \la{d4}
M' = c_0  M \ ,   \ \ \ \qquad 
M\equiv  \h - x^2 + y^2 - p^2 - q^2  + 2  \sqrt{ 1 -\h^2}\,  x p \ , \ \ \ \ \   \ \ \   c_0= - { 4 \h \ov (1 + \h)^2} \ ,   \ee 
where  $c_0$   can  be absorbed into  $\P_0$. 

Thus the   dilaton expected to   be  part of  the  \lmo  target space      background 
in the  ``real"  patch \rf{patch2}  where  the metric  is given by \rf{312} is  in the region \rf{yy} 
 may be written as  (cf. \rf{d2}) 
\be 
e^{\P} = e^{\P_B}\, M   
=  e^{\P_0} \frac{    \h - x^2 + y^2 - p^2 - q^2  + 2  \sqrt{ 1 -\h^2}\,  x p   }
{          \sqrt{-  (1- \h x^2+  \hi y^2)   (1-\h p^2-\hi q^2)} }\ .  \la{d5}  \ee
 Like \rf{d2} this   expression       is  real  if $\P_0$   is real and 
 $1-\h x^2+\hi y^2 \leq 0 , \   1-\h p^2-\hi q^2\geq 0$.\foot{
  The   fermionic   factor  in \rf{d39}  remains real  
  under the analytic continuation (cf. \rf{d39}) 
   with $i$  coming just from  the bosonic 
  square root term $\sqrt{ 1-\h x^2+\hi y^2} $.}

\subsection{RR   background  } \la{secsugra}

The   metric \rf{312} and the  ``bosonic"   dilaton \rf{d2}
were    promoted in \ci{ST}  to an exact type IIB   supergravity solution by supplementing them with an $F_5$   RR field strength  background.  Let  us show  that 
one can   also    find an $F_5$       background   that 
 extends the   metric \rf{312} and the   full dilaton \rf{d5} to 
a  different   10d supergravity solution. 
 The  resulting background   will 
correspond  to   the GS superstring action resulting from the 
$\k$-symmetric \ci{hms2}   \lmo \rf{1} as we   will  show below in section  \ref{next}. 
That    implies  that (in  contrast to   what happens   in the \emo \ci{us3}) 
 the \lmo    represents  
not only a  scale-invariant \ci{holl}, but also  a Weyl-invariant 
sigma  model   and thus defines a consistent superstring theory.

First, let us recall   how  one can  embed
  a  6-torus   compactified 
      background  $M^4 \times T^6$   (e.g. the undeformed \adst  solution  \ci{sor})
  into   type IIB  10d supergravity
(see Appendix A in \ci{lrt}). We shall assume that the $B$-field   and the RR scalar are    vanishing from the start  and 
choose  the  following  ansatz   for the metric and the RR  $F_3$ and $F_5$  field strengths 
($z_i$ are 3 complex coordinates of the 6-torus) 
\be\la{t1}
ds_{10}^2&=&g_{mn}(x) dx^mdx^n+e^{W(x)} dz_id{\bar z}_i\,,
\\
\la{t1a} F_3&= &\tex \frac{1}{2}dC(x)\wedge J_2+\frac{1}{12}\star\big(dC(x)\wedge J_2 \we J_2 \we J_2    \big)\,,  \\
\la{t2} F_5&=&\tex \ha  \big( F \wedge \mbox{Re}\,  \Omega_3 -   
 {F}^* \wedge \mbox{Im} \,  \Omega_3  \big)
%
\ ,  \qquad \qquad  F\equiv \frac{1}{2}F_{mn}(x)dx^m \we dx^n\ ,  \\
J_2 &\equiv&\tex  \frac{i}{2}dz_k\wedge d{\bar z}_k  \ ,\qquad \qquad  \ \  \Omega_3 \equiv  dz_1\we dz_2\we dz_3\,. \la{jjj}
\ee
Here the 4d   scalar $C$  parametrises  the 
 RR 3-form   and $F_{mn}$ is the  4d vector  field   strength (with $F^*$ being its 4d dual) 
 representing the 4d reduction of the RR 5-form. 
To write an  effective 4d action one may  formally relax the self-duality condition  on $F_5$
  replacing  \rf{t2}   with 
$F'_5= {1\ov {\sqrt{2}} } F \wedge \mbox{Re}\,  \Omega_3\,$  which solves the same 
equations of motion (has same stress tensor expressed in terms of $F$). 
 Then the   dimensional reduction of the  relevant bosonic part of the 10d  type IIB  supergravity  action
\be\la{sss1}
S_{10}=\int d^{10} x\sqrt{-g_{10}}\tex \Big[e^{-2\Phi_{10}}(R+4(\del\Phi_{10})^2)-\frac{1}{12}  F_{\m\n\l} F^{\m\n\l} 
-\frac{1}{480}  F_{\m\n\l\r\k} F^{\m\n\l\r\k}\Big]\
\ee
 gives  the following  4d action  ($\Phi\,\equiv  \Phi_{10} -\frac{3}{2}W$)
\be\label{s1}
S=\int d^{4} x\sqrt{-g}\Big[ \tex e^{-2\Phi}\Big(  R+
4( \del\Phi)^2-\frac{3}{2} (\del W)^2  \Big)  -   \frac{1}{4}F_{mn}F^{mn} - 
\frac{1}{8} ( 3 e^W  + e^{-3W})  (\del C)^2\Big]\,.
\ee
This action always   admits a solution with $W=0, \ C=0$   which we  will   assume  in what follows.
The resulting effective  4d action for the metric $g$, dilaton $\P$ and the RR field strength 
 $F_{mn}$  
 becomes simply
 \begin{equation}\label{sact}
{S} = \int d^4x \; \sqrt{-g}\Big[\tex e^{-2\Phi}\big(R + 4(\nabla \Phi)^2\big) - \frac14 F_{mn}F^{mn}\Big]\ .
\end{equation}
The corresponding equations of motion are\foot{Note that   these equations   are invariant under 4d duality rotations of 
the vector  field, i.e.   given a solution   one can construct a  family of solutions 
related by $U(1)$   duality rotations of $F$.  This symmetry is absent  in 10d  where $F_5$ is  self-dual 
(it is compensated by   a rotation of the 6-torus coordinates).}
\be
\label{m1}
 &&R_{mn} + 2 \nabla_m\nabla_n \Phi
= \ha  {e^{2\Phi}}(F_{mp}F_n{}^p - \fo g_{mn}F_{kl} F^{kl}) \ , \\
\la{m2} &&
R + 4 \nabla^2 \Phi - 4 (\nabla\Phi)^2 = 0\ , \qquad \qquad 
\qquad\\
&&  \partial_n (\sqrt{-g}\, F^{mn}) = 0\ , \ \ \ \ \qquad   \del_{[m} F_{nk]}=0 \ .\la{f3}
\ee
As follows   from \rf{m1},\rf{m2}    the dilaton should also satisfy   
\be  \la{m3}  R + 2  \nabla^2 \P=0 \ , \ \ \ \ \ \ \  \qquad    \nabla^2 e^{-2 \Phi} =0 \ . 
\ee
  As  was  found in \ci{lrt} on the example of the  $\eta$-deformed \adst    metric  
  there may be several solutions  for the dilaton and the $F$-form that   solve  \rf{m1},\rf{m2} for the {\it same}
    4d  metric. Thus    given  the  metric  \rf{312}  the solution  for $\P, F$ need not be unique. 
    Indeed, both the  dilatons $\P_B$ in  \rf{d2}   and $\P$ in \rf{d5}   satisfy, as one can check, 
     each of  the two 
    equations in \rf{m3}.\foot{Note that the scalar   curvature for the metric \rf{312}   is 
    $R=  { 4[\hi -(1-\h^2) x^2 ]\ov 1-\h x^2 + \hi y^2}    -   { 4[\hi - (1-\h^2) p^2 ]\ov  1-\h p^2 - \hi q^2}     $.}
    
  Ref. \ci{ST}  found  a real ``bosonic"  solution  
   for $F=dA $   that supports 
    the metric \rf{lag1a} 
    and the  associated   ``bosonic"  dilaton. 
   Written   as a solution   for the  metric 
   \rf{312} and the dilaton \rf{d2} in the algebraic  coordinate patch \rf{yy} 
    the  corresponding vector potential  $A\equiv A_m dx^m$
    takes the following simple form \ci{us1}\foot{Compared to \ci{us1}  here 
    we fix the 4d   vector duality rotation freedom on  the vector potential,
   i.e. choose a particular representative in a class of equivalent solutions related by duality rotations under which the metric and dilaton are invariant. This choice    breaks the formal discrete symmetry  between the 
    coordinates present in the metric \rf{312}.
    Note also   that the factors of string tension $T$  in the metric \rf{312} 
 and the RR field strength  cancel out in the combination 
 $e^{2 \P} F  g^{-1} F  $ appearing in the r.h.s. of \rf{m1} (the l.h.s. also does not depend on $T$). 
 }
   \be \la{m5}
   A_B =  \ha  c\,    \sqrt{ 1 - \h^2}   \   p\,  dy  \ , \qquad \qquad   
  c  =   { 4  \, ( \h\, T)^{-1/2}} \, e^{-\P_0}  \ , 
\ee   
i.e. only one     component  of the field strength  $F_B$  is non-vanishing.
This   solution  becomes imaginary in the coordinate patch \rf{patch1}, i.e. for the metric 
\rf{312}  with $1 - \h x^2 + \hi y^2 >0$  where the dilaton \rf{d2}  
contains an extra factor of  $i$ that can be absorbed into $e^{\P_0}$   so that  it   reappears in 
$A_B$ in \rf{m5} or, equivalently, the RR background $e^\P F$  that supports   the metric \rf{lag1} is imaginary \ci{us1}.


  
The  ``bosonic" solution \rf{312},\rf{d2},\rf{m5} may be  related to  a yet to be  investigated  
alternative  version of  the \lmo  in which one gauges only the bosonic subgroup of $\widehat F$, i.e. where  the gauge fields do not have  fermionic components and thus the dilaton is 
given just by the ``bosonic" one in \rf{d2}.\foot{To construct such a model one may start
with the \adst GS action, split the supergroup element into bosonic and fermionic parts,
 then  write  down the  first-order  action  with respect to the bosonic   currents only, and finally deform this  interpolating action   by replacing the bosonic  analog of the $v$-term in \rf{9}   by the   gWZW model with the  group element $f$    being the bosonic one 
 (the original GS fermions  will thus play  the role of ``spectators" only). 
If such a model will still  represent an integrable
 deformation of the  bosonic non-abelian T-dual of the \adst  model,   the supergroup  invariance (that is apparently 
  broken  by this ``bosonic" construction)   may be recovered at the level of 
a  hidden integrable structure.}

 One may  wonder   why  the background of the  bosonic \lmo should have a \emph{real}  embedding  into supergravity  given that   it is a deformation of a non-abelian T-dual 
 of \adst  background   in all directions  including time. Indeed, it is known that standard abelian T-duality  applied in a time-like direction  maps a real RR background   to an imaginary one 
 (cf. \ci{hull})  and non-abelian T-duality should  be   generalising    the abelian one. 
 However, there is a subtlety in this  argument
which can be understood in our particular case as follows: to be able to apply abelian  T-duality 
 one needs first to take a  limit \ci{us1}  that  ``enhances" the Cartan  directions
 and thus generates  isometries. It turns out \ci{us1}  that  to generate a {\it time-like} isometry 
 one needs to start with the coordinate patch \rf{patch1} , i.e. the metric \rf{lag1}, while 
 taking the limit of the metric \rf{lag1a}   gives a  space-like isometry. Thus  in the case of the metric   \rf{lag1a}   there is actually no reason to expect that   embedding  to supergravity should lead to a  complex solution,  while the RR flux needed  to support the 
  metric \rf{lag1}   is indeed purely imaginary.\foot{We thank B. Hoare for   clarifying discussions  on  this  issue.}   
 
 Like the above ``bosonic" solution of \ci{ST}, 
 our   new solution of \rf{m1},\rf{m2} for $F=dA$  that supports  the metric  \rf{lag1a} 
 or, equivalently,   \rf{312} in the ``physical" region \rf{yy}, 
  and the  
 full \lmo dilaton \rf{d3} or \rf{d5}  turns out to be real.  This is  consistent 
 with the reality property of the  supercoset  \lmo   action \rf{1}. 
 
The    solution   we found 
 is   similar in structure  to the one  discussed 
  in \ci{lrt}   for the $\eta$-deformed  \adst  metric (though   here we have no isometries): 
(i)  the dilaton 
\rf{d5}   contains  a   factor $M=M(x,y,p,q)$  
 that  (in contrast to the metric \rf{312} and the ``bosonic"    $\P_B$ in \rf{d2})
     does not factorise  into  two separate 2d parts,   
and  (ii) the same  function $M$ 
 enters  also  the  vector potential (cf. \rf{m5})\foot{Remarkably, the corresponding $F_{mn}$  still solves   the Maxwell equations in \rf{f3}   despite the fact that  the metric \rf{312} does not contain  any   dependence on  $M$.}
\be \la{m6} 
 A  = \ha   c\   M^{-1}   \, \big[    y\, dx   +       ( \sqrt{ 1 - \h^2}\,     p - x ) \,  dy\big]
   \ , \qquad \qquad    
  c  =   { 4  \, (\h\, T)^{-1/2} }\,   e^{-\P_0}  \ . 
\ee   
 The field  strength $F_{mn}= \del_m A_n -\del_n A_m$ 
that   corresponds to \rf{m6}    has the following  
components\foot{Here 
 $(x^0,x^1,x^2,x^3)= (x,y,p,q)$, i.e. $A_0=\ha c   M^{-1}  y, \ A_1=\ha c   M^{-1}( \sqrt{ 1 - \h^2}\,     p - x  )$.}
\be 
&&  F_{01} =  c\, M^{-2} \h \,  ( 1 - \h p^2 - \hi q^2) \ , \la{ff1}\no  \\
&& F_{02} = c \, M^{-2} \ y\,  ( p - \sqrt{1-\h^2}\, x) \ ,  \qquad \qquad 
F_{03}  =   c  \, M^{-2} \    y \,   q  \ , \no   \\
&& F_{12} = c \,   M^{-2} \ \big[ -  x  p +    \ha   \sqrt{1-\h^2} \, (   \h + x^2 + y^2   + p^2 - q^2) 
 \big]  \ , \no \\ 
&&
F_{13}  =   c \,  M^{-2} \       q\,   (     \sqrt{1-\h^2}\,     p - x)  \ , \qquad \qquad   F_{23}=0 \ .       \la{ff2}
\ee
The  combination that  enters   the   GS superstring  action is   the 
  field  strength  times   the dilaton   with the 
  components taken  in the   vielbein basis (i.e.  $\F_{ab} \equiv  e^\P F_{ab}$), 
  or, equivalently, 
     the  RR bi-spinor  $ \F_{ab} \Gamma^a \Gamma^b$  
     (cf. \rf{31},\rf{3111},\rf{aa1}).  Multiplying \rf{ff2}   by the  dilaton \rf{d5} and the  inverse vielbein factors 
      corresponding to the metric \rf{312} ($g_{mn} =\eta_{ab} E^a_m E^b_n$)
  we find  for $ \F_{ab} $
  \be 
&&  \F_{01} =  ce^{\P_0}\, M^{-1} \h \, \sqrt{-( 1 - \h x^2 +  \hi y^2)  ( 1 - \h p^2 - \hi q^2) }\ ,  \qquad \qquad \F_{ab}   \equiv  e^\P F_{mn}   E^m_a E^n_b \ , \la{fff1}\no  \\
&& \F_{02} = ce^{\P_0}\,   M^{-1} \ y\,  ( p - \sqrt{1-\h^2}\, x) \ ,  \qquad \qquad 
\F_{03}  =   ce^{\P_0}\,  M^{-1} \    y \,   q  \ , \no   \\
&& \F_{12} = ce^{\P_0}\,   M^{-1} \ \big[  - x p + \ha    \sqrt{1-\h^2} \, (   \h + x^2 + y^2   + p^2 - q^2) 
 \big]  \ , \no \\ 
&&
\F_{13}  =   ce^{\P_0}\,  M^{-1} \       q\,   (     \sqrt{1-\h^2}\,     p - x)  \ , \qquad \qquad   \F_{23}=0 \ .       \la{ffff}
\ee
Once again, this   background  \rf{312},\rf{d5},\rf{m6}  is real 
in the region   with $1 - \h x^2 +  \hi y^2 \leq 0$ corresponding to \rf{patch2},\rf{lag1a}.
In Appendix \ref{apa} we will rederive the embedding of the 
solution \rf{312},\rf{d5},\rf{ff2} into   type IIB supergravity directly in 10d. 

As we shall show  in section \ref{next},  this  solution \rf{312},\rf{d5},\rf{m6} 
 is exactly the  background that  appears in the GS  sigma model 
that comes   out of 
 the  supercoset  \lmo  action   \rf{1}  upon integrating out the gauge field $A_\pm$ 
in   the  gauge \rf{patch2}.
  This effectively   demonstrates   that the  \lmo constructed in  \ci{hms2}  
  leads to     a   Weyl-invariant  GS  superstring  action. 

\subsection{Special cases } 

Let us now discuss some  special cases of the solution  \rf{312},\rf{d5},\rf{m6}. 
The first  is $\l=0$ or $\h=1$ (see \rf{219}).\foot{In the \emo this corresponds to the $\vk=i$  point related to the Pohlmeyer reduced theory \ci{hrt,us1}.} 
  In this case  the  metric   \rf{312}  is   that of  a  direct product  of two  2d spaces  corresponding  
 to   the  bosonic  gWZW models $SO(2,1)/SO(1,1)$  and $SO(3)/SO(2)$. It 
 has  two ``rotational"   $SO(1,1) $ and $ SO(2)$   isometries
 (corresponding to   translations in $\xi$ or $\td t$ and $\zeta$ in  \rf{lag1} or \rf{lag1a}). 
 We then get   from \rf{312},\rf{d5},\rf{m6}
   (assuming $1- x^2 + y^2 \leq 0,\ 1- p^2 -q^2 \geq 0$) 
\be 
\h=1\, : \ \ \ && 
\Ti ds^2 = \frac1{1- x^2+ y^2}\big(- dx^2 + dy^2 \big)
+ \frac1{1- p^2- q^2}\big(dp^2 + dq^2\big) \ ,  \la{g1} \\
&& e^\P = e^{\P_0}   {   1 - x^2 +y^2 - p^2 -q^2 \ov \sqrt{ -(1 - x^2 + y^2) (1- p^2 - q^2) }} \  , \\
&&  A=   2 T^{-1/2}  \,   e^{-\P_0}  {1 \ov 1 - x^2 +y^2 - p^2 -q^2 } (    y\,  dx - x\,  dy )  \ .\ee
 In contrast, the ``bosonic"  solution \rf{312},\rf{d2},\rf{m5}   for $\h=1$ becomes 
   just the standard   metric-dilaton gWZW   background  \ci{rab} 
     ($A_B$ in \rf{m5} vanishes for $\h=1$). 
 
Another   special case is    when  $\l=1$   or $\h=0$  and $k$ in \rf{1} sent to infinity
 with  the coordinates $(x,y,p,q)$  and the rescaled tension 
  $h$ in \rf{18} kept fixed.\foot{In the \emo this corresponds to $\vk=0$  or the undeformed 
 \adst 
 theory.}  
 To define this limit we need to start   with the analytic continuation of the solution to the region with 
 $1 - \h x^2 + \hi y^2 \geq 0 , \ \ 1 - \h p^2 - \hi q^2  \leq 0$ 
where the dilaton \rf{d5} is still real. 
       As a result, we get  the following     solution 
\be 
\h=0\, : \ \ \  && 
 h^{-1} ds^2 ={  - dx^2 + dy^2 \ov y^2}
-  {dp^2 + dq^2 \ov q^2 }   \ , \ \qquad \ \  h= \h \, T = \h { k\ov \pi} \ ,  \la{g2} \\
&& e^\P =   e^{\P'_0} \  {  y^2  -q^2 -( x-p)^2 \ov   y \, q } \  , \\
&&  A=   2 h^{-1/2} \,   e^{-\P'_0} \ {1 \ov  y^2  -q^2  -( x-p)^2  } \big[   y\,  dx   - (x-p)\,  dy \big]  \ . \ee
The metric  in \rf{g2} is a  product of $AdS_2$ and $H^2$ (with $--$ signature). In contrast to the 
(analytic continuation of) the standard \adst solution   here it is   
 supported by 
a   non-trivial dilaton and  non-constant 
RR flux.\foot{The isometries of the metric are broken by the $(\P,A)$ background  to just two: 
simultaneous rescaling of all 4 coordinates  and   simultaneous shifts of $x$ and $p$.} 

There is   also  another way   of taking the 
 $\h=0$    limit   (in which the coordinates  $(x,y,p,q)$  are no longer fixed but are scaled in a special way)
 that leads to the
 metric of the  non-abelian T-dual of \adst    (cf. Appendix A in \ci{us1}). 
Consider  first   the second  (``sphere")  part of the metric in \rf{lag1},\rf{311} 
and define 
$z\equiv \cos \vp = \sqrt{ \h p^2 + \hi q^2} \ , \ \   w \equiv \cos \z ={ \h^{1/2} p \ov \sqrt{ \h p^2 + \hi q^2}}$
so that 
$p = \h^{-1/2}   z  w , \ \   q=  \h^{1/2}    z \sqrt{ 1 - w^2} $. 
The standard form of the metric  of the  non-abelian T-dual of $S^2$     is found  by setting 
$z=1 - { \h^2  \ov 2(1 -\h^2)}  Z^2, \ \  w=1 - { \h^2  \ov 2(1 -\h^2)}  W^2$    and taking 
  the limit $\h \to 0$ in \rf{lag1}  for fixed $Z$ and $W$    \ci{klim} 
\be \la{233}
h^{-1} ds^2 =
\,   Z^{-2} ( d W^2 + \fo  [d(Z^2 + W^2)]^2)=   \,     {   dU^2 + d W^2 \ov 2 U - W^2  } \ , 
\qquad \ \ \  U\equiv  \ha (Z^2 + W^2)\ ,  \ee 
where,  as in \rf{g2},    the rescaled  tension $h$ is   again  fixed in the limit. 
   The  coordinates $(p,q)$ used  in \rf{312}   are thus no longer fixed   in 
 this $\h \to 0$  limit. 
 Let us set 
 \be \la{224}  x=  \h^{-1/2} (1   +  \h^2  V), \ \qquad   y= \h^{3/2} Y\ , 
  \quad 
 \qquad p =  \h^{-1/2}  (1   -   \h^2  U)\   ,\qquad
  q= \h^{3/2} W  \ , \ee
  where  to satisfy the ``physical" patch \rf{lag1a} conditions 
   $1 - \h x^2 + \hi y^2 <0, 1 - \h p^2 - \hi q^2>0$ 
  we assume   that $ Y^2 < 2 V$, $W^2 < 2 U$. 
  Then 
 the $\h\to 0, \ h=$fixed  limit of the   metric \rf{312}  with fixed  new coordinates $(V,Y,U,W)$   
becomes the metric of the  non-abelian T-dual of \adst 
\be \la{1122} 
h^{-1} ds^2 = 
 { d V^2 - dY^2 \ov 2 V - Y^2 } +       {   dU^2 + d W^2 \ov 2 U - W^2  }   \ . \ee
 $\P$ in \rf{d5} and $A$ in \rf{m6}     take the following form 
\be \la{22a} 
e^{\P} = e^{\P'_0} {    2 (U-V )-1  \ov   \sqrt{ (2 V - Y^2) ( 2 U - W^2)  }} \ , \qquad\qquad 
A=  2   h^{1/2}   e^{-\P'_0} {1 \ov     2 (U-V) -1}     Y d V \ , 
\ee
where  in $A$ we have  dropped   a pure gauge term $\sim dY$.\foot{ 
Note that in this  limit  the ``bosonic" solution of \ci{ST} 
leads to the same metric \rf{1122}  supported by a different   combination of  fields: 
$e^{\P_B} = {e^{\P'_0}   \ov  \sqrt{ (2 V - Y^2) ( 2 U - W^2)  }}, \ \ 
A_B= 2 h^{1/2}   e^{-\P'_0}     U d Y. 
$
}

\subsection{Scaling limits} 


As  was  found   in \ci{us1},  making 
 the formal  coordinate redefinition 
$(t,\xi;\vp,\zeta) \to (t,\rho; \vp, r)$  in \rf{lag1}  
combined with infinite imaginary shifts of the
$(t,\vp)$ directions    and setting $\vk= i \h$ 
\begin{align}
&  t\to t+{\tex\frac{i}{2}} \log\frac{1-\varkappa^2\rho ^2}{1+ \rho ^2}+ i \log \g_1\ , &&
\xi \to
\ha \log \frac{-1+\varkappa \rho }{1+\varkappa \rho } \ , \nonumber
\\&\varphi \to \varphi+  {\tex\frac{i}{2}}   \log \frac{1+\varkappa^2r^2}{1 -r^2} + i \log \g_2\ , &&
\zeta \to {\tex \frac{i}{2}}
\log \frac{1+i\varkappa r}{-1+i\varkappa r} \ , && \g_1,\g_2 \to \infty \ .\label{part2}
\end{align}
  transforms  the metric \rf{lag1} into
\begin{equation}\begin{split}\label{metdmv2}
 h^{-1} ds^2 = \,  \frac{1}{1-\varkappa^2\rho^2}\big[-(1+\rho^2)dt^2 + \frac{d\rho^2}{1+\rho^2}\big]
+ \frac{1}{1+\varkappa^2 r^2}\big[(1-r^2)d\varphi^2 + \frac{dr^2}{1-r^2}\big] \ , 
\end{split}\end{equation}
which is   the $\eta$-deformed \adst  metric
\ci{dmv,abf1,hrt,Fateev:1992tk}
with $h\equiv  {k\ov \pi} \h$ as  string tension.
This  limiting   procedure becomes   more transparent  
in the algebraic  coordinates \rf{312}.
Here we will consider the patch  where $ 1- \h x^2 + \hi y^2 >0, \ 1-\h p^2 - \hi q^2   <0$. 
Performing   independent  infinite   rescalings of  the  coordinates $(x,y)$ and $(p,q)$ 
\be 
(x,y) \to \g_1 (x,y) \ , \ \ \ \ \ \  (p,q) \to \g_2 (p,q) \ , \  \ \ \ \ \qquad \g_1,\g_2 \to \infty \ \la{322}
\ee
generates   scaling  isometries  in each of the  two factors of the metric \rf{312}:
\begin{equation}\la{3122}
 ds^2 = \frac1{-\h x^2+\hi y^2}\big(- dx^2 + dy^2 \big)
+ \frac1{-\h p^2-\hi q^2}\big(dp^2 + dq^2\big) \ .
\end{equation}
Doing analytic continuation of coordinates, setting $\h= - i \vk$ (cf. \rf{219})  
and absorbing the overall  factor of $\h$ into $h$ in \rf{18} converts  this metric 
into \ci{us1}
\begin{equation}\la{316}
 h^{-1}  ds^2 =   \frac1 {y^2 - \varkappa^2x^2 }\big( dy^2 + dx^2 \big)
+ \frac 1 {q^2 + \varkappa^2p^2}\big( -dq^2 + dp^2 \big)  \ .
\end{equation}
This  may be interpreted as the metric of $\eta$-deformed $H^2 \times dS_2$
background which is related to $AdS_2
\times S^2$ by analytic continuation.
In this  scaling limit \rf{322}  (combined with a shift  of the  constant part of the dilaton) 
  the  ``bosonic"  solution \rf{312},\rf{d2},\rf{m5} 
thus reduces to the metric   \rf{3122} and 
  \be 
e^{\P_B} =  { e^{\P'_0} \ov \sqrt{ (\h x^2-\hi y^2)   (-\h p^2-\hi q^2)} }\  , 
\qquad \qquad      A_B= \ha c\, e^{-\P'_0}    \sqrt{ 1 - \h^2} \,   p\,  dy  \ . \la{57}  \ee
As was shown in \ci{us1,us2},   written in angular coordinates and after  an analytic continuation 
the metric \rf{3122} and $e^{\P_B}  F_B$ (but not the  dilaton, cf. \ci{us3}) of this background   is  found to  be 
T-dual  to those of the   $\eta$-deformed \adst   background.

In the case of 
 the  new solution \rf{312},\rf{d5},\rf{m6} 
the  infinite rescaling \rf{322} with fixed  $ {\g_1\ov \g_2}$ 
gives  again 
the metric  \rf{3122}  while the dilaton   and the RR gauge field     become
\be 
&&e^{\P} = \frac{  e^{\P_0}\  \td M }
{          \sqrt{ (\h x^2- \hi y^2)   (-\h p^2-\hi q^2)} }\ , \ \ \ \la{h1} \\ 
&& \td M \equiv   (\g_1 \g_2)^{-1}  M 
=   - {\g_2\ov \g_1} (p^2 +q^2)   +  {\g_1\ov \g_2}  (  - x^2 + y^2)     + 2   \sqrt{ 1 -\h^2}\,  x p \ , 
   \la{d55} \\
&&A  =   \ha c\, e^{-\P_0}   {\td M}^{-1}   \, \big[  \sqrt{ 1 - \h^2}\,     p\,   dy     +   {\g_1\ov \g_2}     (y\, dx  - x\, dy)
\big]
   \ .  \la{m69} 
\ee   
 While the metric  \rf{3122}   has two scaling  isometries,   
     $\P$ and $A$  have  only one isometry under the same 
     simultaneous rescaling of all the coordinates.\foot{In contrast, the  bosonic dilaton  in \rf{57} 
     is invariant only under the opposite scaling   of $(x,y)$   and $(p,q)$.}
     This limiting background  \rf{h1},\rf{m69} 
  (which  is obviously different from \rf{57}) 
  is of course  still a solution of  the supergravity equations \rf{m1},\rf{m2} 
and thus defines a consistent GS sigma model  
which is   classically T-dual to the    GS action   for the   $\eta$-deformed \adst model. 
 Absence  of the two isometries  in the dilaton  does  not allow one   to perform the T-duality on the whole background
 (i.e. at the quantum level).\foot{Let  us note    that this   solution   is  also different from a  class of solutions  with $\eta$-deformed \adst metric
  found in \ci{lrt} as there the existence of two  separate $U(1)$ isometries   was assumed  from the start.}

Let us now consider a particular ``asymmetric"    case   of the  scaling limit \rf{322} 
 with $  {\g_1\ov \g_2}  \to 0$. 
Then   $\td M  \to   - {\g_2\ov \g_1} (p^2 +q^2)  $  and     \rf{h1},\rf{m69}   become\foot{Note that 
   the alternative limit ${\g_1\ov \g_2}  \to \infty $
   leads to a   different  background  that  has to do with our particular choice  of the vector potential 
on the orbit of duality transformations that breaks symmetry between $(x,y)$  and $(p,q)$ pairs of coordinates.}
 \iffa  
gives 
$e^{\P} = \frac{  e^{\P'_0}\   (x^2 -y^2)  }
{          \sqrt{ (\h x^2-\hi y^2)   (-\h p^2-\hi q^2)} }$ with $e^{\P'_0} \equiv -  {\g_1\ov \g_2} e^{\P'_0}
$ and $A  =  \ha  c\, e^{-\P'_0}   {(x^2 - y^2 ) }^{-1} (y dx - x dy)  $.
But then  $A$ is pure-gauge,  while  after $x'= {x \ov x^2 - y^2}, \ y'= {y \ov x^2 - y^2}$
the metric and dilaton take ``bosonic" limit form  so how then eqs of motion are solved???
The fact that this limit looks  different has to do with our particular choice  of the vector potential 
on the orbit of duality transformations that breaks symmetry between $(x,y)$  and $(p,q)$ pairs of coordinates. ????
}
\fi 
\be 
&&e^{\P} = \frac{  e^{\P'_0}\   (p^2 + q^2)  }
{          \sqrt{ (-\h x^2+\hi y^2)   (-\h p^2-\hi q^2)} }\ , \ \ \ \  \qquad  e^{\P'_0} \equiv -  {\g_2\ov \g_1} e^{\P'_0} \ ,    \la{89}
  \\ 
&&A  =  \ha  c\, e^{-\P'_0}   {(p^2 + q^2) }^{-1}   \  \sqrt{ 1 - \h^2}\,     p\,   dy    
   \ .  \la{99} 
\ee   
The fact that the dilaton becomes factorizable  and that $A$  looks similar to $A_B$ in \rf{57} 
is not   accidental. 
Doing the coordinate  transformation $(p,q) \to (p',q')$
\be \la{bh}
p' = { p\ov p^2 + q^2} \ , \ \ \qquad \qquad  \ \ \  q' = { q\ov p^2 + q^2} \ , \ee 
that preserves the form of the metric  \rf{3122}
we   find that   \rf{3122},\rf{89},\rf{99} take the  same   simple  form  as in \rf{57}
\be\la{3338}
 &&ds^2 = \frac1{-\h x^2+\hi y^2}\big(- dx^2 + dy^2 \big)
+ \frac1{-\h p'^2-\hi q'^2}\big(dp'^2 + dq'^2\big) \ , \\
&&e^{\P} = \frac{  e^{\P'_0} }
{          \sqrt{ (\h x^2-\hi y^2)   (-\h p'^2-\hi q'^2)}     }\ , \qquad \quad 
A  =  \ha  c\, e^{-\P'_0}   \  \sqrt{ 1 - \h^2}\,     p'\,   dy    
   \ .  \la{999} 
\ee   
Thus   the  ``asymmetric" scaling limit
  \rf{322}    
with  $\g_1,\g_2\to \infty$  and 
${\g_1\ov \g_2}  \to 0$ of   the  \lmo  background  \rf{312},\rf{d5},\rf{m6}     is equivalent to 
the scaling limit \rf{57}  of the  ``bosonic"  solution of \ci{ST}.\foot{For the dilaton \rf{d5}  that was already 
pointed out in \ci{us1}. We thank   B. Hoare   for   discussions of this limit 
and suggesting the existence of the  transformation \rf{bh}  that demonstrates also the equivalence of the 
limiting RR backgrounds.
} 

We conclude that   the  proposal   of \ci{us1}
 that a 
   scaling limit  of the \lmo   should  give a background   which is classically T-dual to the \emo   background 
  is now  confirmed    for the RR background as well. 
The  underlying reason  why the  above ``asymmetric" scaling limit  of the \lmo   is required to recover the T-dual 
of the \emo  and why this limit is the same as the limit of the ``bosonic" solution 
should  be  related to  the fact that such a limit ``enhances" the {\it bosonic}    Cartan directions 
suppressing  the  effect of  gauging of  the fermionic directions   and thus 
ameliorating the distinction between   gauging the full  supergroup  in the \lmo  and just its bosonic 
part as in the model that should   correspond to the ``bosonic" background of \ci{ST}.

\section{RR   background   from supercoset \lmo }
\label{next} 

Let us   now show that the RR background \rf{t2},\rf{ff2}    appears   in  the GS superstring   sigma model 
that   emerges from  the \lmo   action \rf{1}  upon  integrating out  the gauge fields $A_\pm$. 
As was  shown in \ci{hms2}, the \lmo   action  has a local fermionic  symmetry that  may be
 interpreted as  $\k$-symmetry of the resulting GS action. Expressing  the supergroup field  $f$ 
  in terms of the bosonic and fermionic coordinates 
 and expanding to quadratic  order in fermions one  may  be able to put 
    the resulting  quadratic  fermionic action   into the standard type IIB  GS form 
\be\label{31}
\begin{aligned}
S_{2}&=  \int \, {\rm d}\sigma {\rm d} \tau\ i\, \bar{\Theta}_I \Pi_+^{IJ\a\b} {E}^a_\a \G_a \, {D}^{JM}_\b \Theta_M\,,
\qquad \tex \qquad 
\Pi_\pm^{IJ\a\b}=\frac{1}{2}\left( \g^{\a\b} \delta^{IJ} \pm \epsilon^{\a\b} \sigma_3^{IJ} \right)\,,
\end{aligned}
\ee
where $\Theta_1,\Theta_2$ are $32$-component Majorana-Weyl spinors of positive chirality ($\G_{11}\Theta_I=\Theta_I$), 
 ${E}^a_\a= E^a_\m \del_\a X^\m $ is the pullback  of the vielbein 
for  the  metric  $g_{\m\n}=\eta_{ab}E_\m^a E_\n^b$ in  the bosonic  part of the $\sigma$-model,  
 and  $\g^{\a\b}\equiv \sqrt { - h}\,  h^{\a\b}$ where $h^{\a\b} $ is the world-sheet metric.\foot{Here  $\a,\b=0,1$   and  $\m,\n, ...$  and $a,b,...$ are 10d   coordinate and 
 tangent space indices respectively. $I,J,M,...=1,2$   will denote the labels of  the two MW spinors.}  
For a generic IIB  supergravity  background, the operator $D^{IJ}_\a$  in \rf{31}  takes the form
(see, e.g., \cite{Cvetic:1999zs,Wulff:2013kga})
\be\la{3111}
\begin{aligned}
{D}^{IJ}_\a  = &\tex 
\delta^{IJ} \big( \pa_\a  -\frac{1}{4} {\omega}^{ab}_\a \G_{ab}  \big)
+\frac{1}{8} \sigma_3^{IJ} {E}^a_\a {H}_{abc} \G^{bc}
\\
&\tex -\frac{1}{8} e^{\Phi} \left( \epsilon^{IJ} \G^a {F}_a + \frac{1}{3!}\sigma_1^{IJ} \G^{abc} {F}_{abc} + \frac{1}{2\cdot5!}\epsilon^{IJ} \G^{abcde} {F}_{abcde}   \right) {E}^h_\a \G_h.
\end{aligned}
\ee
Assuming that one finds  the action in the form 
 \rf{31}  one   can then extract   the combination $\F= e^\P F$ of the RR field strengths and dilaton 
from the the operator $D^{IJ}_\a$ (cf. \ci{abf2} in the case of the \emo). 
 In the present  case of the \lmo   there is a  natural  candidate \ci{hms2}  for the dilaton expressed in terms of the 
  superdeterminant of the matrix  in the  quadratic $A_+ A_-$ term in \rf{1};  that  should   allow  to extract the RR   flux $F$ itself
  ($F$ should then   satisfy the Bianchi  identities  if the whole construction is consistent).  

For  the standard GS action in type IIB   supergravity background \ci{gris} 
 the sum of the bosonic  and   quadratic   fermionic   action 
  is invariant (to leading order in  $\Theta$) under  the  $\k$-symmetry variations for the $\sigma$-model fields
  and the  2d metric (here $K_{\a I}$  are the Grassmann 32-component spinor $\k$-symmetry parameters)
\be
&& \delta X^\m  \tex =  \frac{i}{2} \ \bar{\T}_I \G^\m  \delta  \T_I + \mathcal{O}(\T^3)\,,
\qquad \qquad \G^\m={E}^{\m a}  \G_a\,, \label{32}
\\
&&\delta  \T_I \tex = \frac{1}{2} \Pi_-^{IJ\a\b}   \G_\b  {K}_{\a J}+ \mathcal{O}(\T^2)\,,
\qquad \qquad \G_\b={E}_{\b}^a  \G_a\,, \label{323}\\
&&\delta  \g^{\a\b}=
- 2i\ \Pi_+^{IJ\, \a\a'}\Pi_+^{JN\, \b\b'}
\ \bar{{K}}_{I\a'}{D}^{NL}_{\b'}\Theta_{L}+ \mathcal{O}(\T^3).   
\la{333}
\ee
Thus  if  $\k$-symmetry is in place, we may then extract  the 
 operator ${D}^{IJ}_\a$ containing the information  about  the background RR fields  not from the   action \rf{31} 
 directly
 but  rather from the expected form of the $\k$-symmetry  variation of the world-sheet metric in \rf{333} 
  which is \emph{linear} in $\Theta_I$.\foot{Deriving the  explicit  form of the quadratic  fermionic action from the \lmo 
  and comparing to the expected form~\eqref{31} is an involved calculation. 
  One difficulty lies  in the fact that a  random  choice of  the bosonic and fermionic coordinates $(X^\m,\Theta_I)$
   is likely not be the right one to get the action in the standard   GS form \rf{31}, i.e. 
    one would  need to find a proper field redefinition   to match \eqref{31}.
In general,   this  would involve   rotating  the fermions by an ($X^\m$-dependent) matrix, and also shifting the 
 bosons by terms  quadratic in $\Theta_I$,     generating extra $\Theta^2$   terms from  the  change  of the 
  bosonic action (see below).} 
  
  Let us note that a   generic choice of the 
  coordinates $(X^\m,\Theta_I)$ in the \lmo     action 
  will not   lead to the  \emph{standard}  GS  form \rf{32}--\rf{333}  of the $\k$-symmetry variation for the world-sheet metric. 
  A natural way to find the right coordinates is to study the $\k$-symmetry variations of $(X^\m,\Theta_I)$  first 
  and find the proper field redefinition that puts them into  the form~\eqref{32}--\rf{333}.
This  will be the strategy that we will  use  here.  


The invariance of the \lmo  \rf{1}  under   the fermionic   symmetry 
 was proved in~\cite{hms2} in the conformal gauge (i.e.  using the Virasoro constraints). 
The action was found to be  invariant under the two independent variations $\delta_1,\delta_2$ of $f$ defined by 
\be\label{34}
\begin{aligned}
\op_+^{-1}(\CF^{-1}\delta_1\CF)=A_-^{(2)}\tilde\alpha+\tilde\alpha A_-^{(2)}\ ,\qquad\qquad \tilde\alpha\in\hat\mf^{(1)}\,,\\
\op_+^{-1}(\CF^{-1}\delta_2\CF)=A_+^{(2)}\hat\alpha+\hat\alpha A_+^{(2)}\,,\qquad\qquad  \hat\alpha\in \hat{\mathfrak f}^{(3)}\,,
\end{aligned}
\ee
where one needs to   substitute   the solutions for the gauge fields in \rf{1}
\be\label{35}
A_\pm=\mp\op_\pm^{-1}(\CF^{-1}\partial_\pm\CF)\ 
\ee
projected to the coset part of the superalgebra. 
Here   the linear operators $\op_\pm$ act on a generic element $\M$ of the superalgebra  as
\be\label{36}
\op_+(\M)=\CF^{-1}\M\CF -\Omega^T(\M)\,,
\qquad\qquad 
\op_-(\M)=\M-\CF^{-1}\Omega(\M)\CF,
\ee
where $f$ is an element of the supergroup $\hat F$ and (cf. \rf{pii})
\be\la{37}
\Omega= P^{(0)}+\lambda^{-1}  P^{(1)}+\lambda^{-2} P^{(2)}+\lambda P^{(3)}\ ,\quad \qquad 
\Omega^T= P^{(0)}+\lambda P^{(1)}+{\lambda^{-2} }P^{(2)}+\lambda^{-1} P^{(3)}\ .
\ee
While in \ci{hms2} where  the conformal gauge was assumed
 the 2d metric was  not transformed and instead the Virasoro constraints were used, 
we can also deduce  a ``conformal-gauge'' version   $\delta \gamma |_{c.g.}$ of the $\k$-symmetry
 variation of the 2d  metric
  obtained by formally imposing  the conformal gauge on the r.h.s. 
  of~\eqref{333}.\foot{It would be interesting to  repeat the derivation of~\cite{hms2} without imposing the 
  conformal gauge, but for us $\delta \gamma^{\a\b}  |_{c.g.}$ will be enough in order to extract the operator $D^{IJ}_\a$ unambiguously.}
   The two independent $\k$-symmetry variations that we will need to compute are~\cite{hms2}
\be\label{310}
\begin{aligned}
\delta_{1}\gamma^{--} \big|_{c.g.} = -2\lambda^2\ \STr\big(W[\tilde\alpha,A_+^{(1)}]\big)\,,\qquad \qquad 
\delta_{2}\gamma^{++} \big|_{c.g.} =-2\lambda\ \, \,\STr\big(W[\hat\alpha,A_-^{(3)}]\big)\,,
\end{aligned}
\ee
where $W$ is defined in \rf{b3}. 
This  will allow us to extract the RR field  background. 

A crucial  step in the derivation  turns out to  be  the inversion of the operators $\op_\pm$ in \rf{36} as their 
action is quite involved, especially on the odd part of the superalgebra.
Here  we will consider     only the   simplest case    of the \lmo   for the 
   \adst supercoset. 
   We shall  use the explicit   representation   of $\su(1,1|2)$ superalgebra in terms of 
    $4\times 4$ matrices  described   in Appendix \ref{apb}.


\subsection{Choice of group element   and  $\k$-symmetry variations of coordinates}

We shall choose the  gauge-fixed  group element $f \in PSU(1,1|2)$
 as   a product of an element $f_B$  corresponding to  the bosonic subalgebra,  and  the   fermionic   part $f_F$
 \be\label{3200}
f=f_B\,  f_F \ , \qquad \qquad f_B=\check f_B \oplus  \hat f_B \ ,\qquad  \qquad f_F=\exp(\gen{Q}^I \theta_I) \ , 
\ee
where we  found it convenient to   choose   
\be\la{3210}
\check{f}_B=e^{\frac{1}{2}\xi \sigma_1}e^{it \sigma_3}e^{\frac{1}{2}\xi \sigma_1}\,,
\qquad\qquad
\hat{f}_B=e^{\frac{i}{2}\zeta \sigma_1}e^{i\varphi \sigma_3}e^{\frac{i}{2}\zeta \sigma_1}\,.
\ee
This  parametrisation  of bosonic coordinates 
  is related to \rf{patch1}  by  a gauge transformation. The  expressions in the ``real"  patch \rf{patch2} 
may be  obtained by a simple analytic   continuation. 
 While    in the case of the \adst supercoset action     the choice~\eqref{3200} for the group
  element\footnote{In the supercoset construction, the choice~\eqref{3200} should be accompanied by a 
  proper parameterisation of the group element, compatible with gauge transformations that act only from the right.}  
  directly  leads to the standard   GS type    quadratic fermionic action, in the \lmo case 
one    would need  an additional $\l$-dependent redefinition  of the fermionic and bosonic coordinates  in order to put the action in the GS form.


The bosonic part of the action of the $\lambda$-model   contains  the  
 metric $g_{mn}$ \rf{lag1},\rf{312} and no $B$-field~\cite{ST,hms2}.
There are two natural equivalent ways to define
a vielbein $E^a= E^a_m dX^m$  corresponding to  $g_{mn}$
\be\la{3222}
\begin{aligned}
{E}^{(\pm)a} &=2 \cla\,  \STr\left[ \gen{P}^a (\op^{(0)}_\pm)^{-1}( \CF^{-1} d \CF)\right],
\qquad\qquad \tex 
\cla =\frac{\sqrt{1-\lambda ^4}}{2 \lambda ^2}= { \sqrt \h \ov 1- \h } \ , 
\end{aligned}
\ee
where the superscript $(\pm)$ on $E^a$ indicates  which linear operator $\op^{(0)}_\pm$  is entering its definition.
The superscript ${(0)}$ means that we switch off the fermions in $\op_\pm$ in \rf{36}
(see Appendix \ref{apb}  for notation). 
Then using  \rf{3200},\rf{3210}   we  find  (see Appendix~\ref{apc} for the details)
\be\label{3237}
\begin{aligned}
&\tex {E}^{(\pm)0} =\pm\frac{2 \lambda ^2 \cla }{\lambda ^2-1}(\cosh \xi \,  dt-  \sinh \xi  \cot t\, d\xi)\ ,\quad
&&\tex {E}^{(\pm)1} =\frac{2 \lambda ^2 \cla }{\lambda ^2+1}(\sinh \xi  \,   dt- \cosh \xi  \cot t\, d\xi )\ ,\\
&\tex {E}^{(\pm)2} =\pm\frac{2 \lambda ^2 \cla }{\lambda ^2-1}(\cos \zeta\,   d\varphi + \sin \zeta  \cot   \varphi\, d\zeta  )\ ,\quad
&&\tex {E}^{(\pm)3} =\frac{2 \lambda ^2 \cla }{\lambda   ^2+1}(\sin \zeta \,  d\varphi -  \cos \zeta  \cot \varphi\, d\zeta )\ ,
\end{aligned}
\ee
so that  ${E}^{(+)a}$   and   ${E}^{(-)a}$  are related by a Lorentz transformation\footnote{This Lorentz transformation has a 
 simple form because of the gauge choice for the bosonic group element \rf{3210}. For a generic gauge choice the Lorentz transformation will depend also on the  bosonic coordinates.}
\be\label{324}
{E}^{(-)a}=\Lambda^a_{\ b}{E}^{(+)b}\,,
\qquad\qquad 
\Lambda^a_{\ b}=
\text{diag}\left(-1,1,-1,1\right)\,.
\ee
The corresponding metric is the one in \rf{lag1}  (in this section we set tension $T=1$). 
In the algebraic coordinates \rf{1z},\rf{311}  where the metric is  given by \rf{312} 
the ${E}^{(\pm)a} $ take the obvious diagonal form.
Below we will  use the $E^a={E}^{(+)a}$   choice. 

Let us now  turn to   the $\k$-symmetry variations of bosonic and fermionic coordinates found 
 by computing~\eqref{34} explicitly in  the   parameterisation  \rf{3200},\rf{3210}   (see Appendix~\ref{apc}).
When we project~\eqref{34} on odd generators and  expand to leading order in fermions, 
we find   the $\k$-symmetry transformation for the fermions $\delta\theta_I$ in terms of 
the corresponding  parameter $\vka_I$ defined in \rf{kappaI}.  To put it into  the standard GS form  we need  to  redefine  $\theta_I$ and  the 
 parameters $\vka_I$ as
 \be\label{327}
\theta \to  \cla ^{-1/2}\, k_{+,F}^t \cdot (\lambda^{-1}U\oplus \mathbf{1}_4)\,  \theta,
\qquad  \qquad  \vkappa \to 2 \cla ^{1/2}\left[ (-\lambda^{-1}U)\oplus \mathbf{1}_4 \right]\vkappa\ ,
\ee 
where 
 we have collected  $\theta_I$ and $\vkappa_I$ into the 2-vectors $\theta$
 and $\vkappa$. 
Here $k_{+,F}$ is the $8\times 8$ matrix which encodes the action of the operator $\op^{(0)}_+$ on the odd generators of the algebra (see eq.~\eqref{c8}), while $U$ is the $4\times 4$ matrix (acting only on spinor indices of $\theta_1$ and $\vkappa_1$) which implements the Lorentz transformation $\Lambda$ in \rf{324}  on the fermions. 
Writing the result   in 10d notation  we get for the non-vanishing $\k$-symmetry variations\footnote{These transformations 
 should be compared to~\eqref{323} after we fix conformal gauge. We follow the conventions of~\cite{hms2} and take  $\sigma^\pm= \tau\pm \sigma$ and $\gamma^{\tau\tau}=-\gamma^{\sigma\sigma}=\epsilon^{\tau\sigma}=-\epsilon^{\sigma\tau}=1$, so that $\gamma^{+-}=\gamma^{-+}=\epsilon^{-+}=-\epsilon^{+-}=2$. It is assumed that here  the index $a$ runs only from 0 to 3.}
\be\la{328}
\delta\Theta_1 &=  {E}^{a}_-\G_{a}\, K_{1}\,,
\ \qquad \qquad 
\delta\Theta_2 & = {E}^{a}_+\G_{a}\, K_{2}\,.
\ee
Here $\G_a$ are   $32 \times 32$ gamma matrices (defined  in Appendix~\ref{apd}), 
and we embedded the 4-component spinors $\theta_I$ and $\vka_I$ into  the 32-component spinors $\T_I$ and $K_I$ in \rf{323} 
as
\be
\la{329}
\tiny \T_I = \left(\begin{array}{c} 1 \\ 0 \end{array}\right) \otimes \theta_I \otimes \left(\begin{array}{c} 1 \\ 0 \\ 0 \\ 0 \end{array}\right)\,,\qquad\qquad
K_I = \left(\begin{array}{c} 0 \\ 1 \end{array}\right) \otimes \vka_I \otimes \left(\begin{array}{c} 1 \\ 0 \\ 0 \\ 0 \end{array}\right)\,,
\ee
so that $\G_{11}\Theta_I=+\T_I$ and $\G_{11}K_I=-K_I$.

Next, if  we project~\eqref{34} on the  bosonic coset generators of  the superalgebra and keep only the leading order term  
 in fermions, we find an equation for the $\k$-symmetry variation of the bosonic coordinates $\delta X^m$.
To put it into the standard form \rf{32}, i.e.  
\be\tex 
 \delta X^m =  \frac{i}{2} \ \bar{\T}_I \G^m  \delta\T_I \,,\qquad \qquad 
\qquad \G_m={E}^{a}_m  \G_a\,, \la{3301}
\ee
 we need, in addition to  the 
   redefinition of    the fermions ~\eqref{327},
    to do a  redefinition of the bosons\foot{Note 
   that an  attempt to  put $\delta_{}X^m$ into the standard form by shifting the bosons as in~\eqref{3319}   could, in principle, 
    fail.  Indeed, $\Delta_{IJ}^m$ must  have  a definite symmetry property in order for  the shift of $X^m$  not to 
    vanish because of the Grassmann nature of $\Theta_I$.
If the terms that we want to cancel  by redefinition do not have the same 
symmetry property, then~\eqref{3319} is not enough to get to the standard form.} 
\be\label{3319}
X^m\to X^m + \bar\T_I \Delta_{IJ}^m \T_J\, , 
\ee
where $\Delta_{IJ}^m $ is defined in \rf{c12}.    
It should be noted   that  for  our present  purpose   of extracting the  RR  background, i.e. the information   about the derivative \rf{3111} 
from \rf{333}, 
 the study of the $\k$-symmetry variations of the  bosons is not required    and  is just a check.
  In fact, the redefinition \rf{3319}  does not modify the result for the $\k$-variation of the world-sheet metric  \rf{310},\rf{333}
   at the  leading order in fermions. However,  one would need to know the explicit form of \rf{3319}  in order to   find the RR  background directly from the  action as that would require   putting  the quadratic   fermionic term  in the \lmo action into the standard GS form \rf{31}.

It is worth mentioning that our results prove that the action at quadratic order in fermions will be of the standard form of GS. In fact, the full action is invariant under a local fermionic symmetry which we found to be the standard $\k$-symmetry at the leading order in fermions. 

\subsection{RR   background  from $\k$-symmetry variation of the  world-sheet metric}

Let us now  consider the variation of the   world-sheet metric. 
Starting   with 
~\eqref{310} we  first  need to compute $A^{(1)}_+$ and $A^{(3)}_-$ at linear order in fermions,  do the field redefinitions~\eqref{327}
and then 
 compare to~\eqref{333}, where one formally needs  to  impose  the  conformal gauge on the right-hand side
\be\la{3325}
\begin{aligned}
\delta\g^{--}\big|_{c.g.}= -8i\ \bar{{K}}_{1}{D}^{1J}_{+}\Theta_{J}\,,
\qquad \qquad 
\delta\g^{++}\big|_{c.g.}= -8i\ \bar{{K}}_{2}{D}^{2J}_{-}\Theta_{J}\,.
\end{aligned}
\ee
We   then isolate the  contributions  in $D^{IJ}$ 
depending on the tensors $\delta^{IJ},\sigma_1^{IJ},\epsilon^{IJ},\sigma_3^{IJ}$    and compare to \rf{3111}.
The   $\delta^{IJ}$   structures are found to   be given by   the standard  derivatives of  the fermions  (with no 
 unwanted matrix rotation of  the spinor indices) plus  terms with the 
  spin connection $\omega^{ab}$ constructed from the vielbein $E^a=E^{(+)a}$ in~\eqref{3237}, in agreement with \rf{3111}. 
  Another consistency check 
 is the absence of a contribution proportional to $\sigma_3^{IJ}$ reflecting  the vanishing  of the $H$-field  background
 (absent in  the bosonic part of the action \ci{hms2,ST}).\foot{Knowing 
 that  the $\k$-symmetry variations of both the bosonic and fermionic coordinates are put  in the 
 standard form, it is   then 
not  surprising that we find the expected values of  $\omega^{ab}$ and $H$:
 the action is invariant under $\k$-symmetry, and at leading order this symmetry is indeed 
 relating $\omega^{ab}$ and $H$ to the metric $g_{mn}$ and the  $B_{mn}$ field  appearing in the bosonic action.}.
Thus  from ~\eqref{310} we get
\be\la{3337}\tex
D^{IJ}_\a= \delta^{IJ}\big(\partial_\a -\frac{1}{4}{\omega}^{ab}_\a\Gamma_{ab} \big)+ \frac18\SS^{IJ} {E}^a \Gamma_a\ , 
\ee
where $\SS^{IJ}$ is off-diagonal in  $I,J$ and   should thus represent  the contribution of  the RR fields in \rf{3111}. 
Since we can compute the $\k$-symmetry variations $\delta_1$ and $\delta_2$ independently, we can easily check that 
$\SS^{IJ}$  is proportional to  $\epsilon^{IJ}$. 
The   absence of  $\sigma_1^{IJ}$ term implies, according to \rf{3111}, the    vanishing of the 
RR 3-form.\footnote{This   argument   illustrates  the power of our method of extracting the RR background 
from the structure of the $\k$-symmetry transformations: 
 if one   would  try   to extract this information 
from the computation of the action one  would first need to project $\SS^{IJ}$ on the product of  3 gamma matrices. 
The  central point is  that   the action is quadratic in the same $\T_I$, 
while the $\k$-symmetry variation of the world-sheet metric 
contains  two different Grassmann spinors $\T_I$ and $K_I$.}

Next,   let us     make an assumption  that also the RR 1-form term  in \rf{3111} vanishes,  i.e. that we should    have 
\be\tex \la{3341}
\SS^{IJ}=-  \frac{1}{2\cdot 5!}\,  \epsilon^{IJ} \, e^{\Phi}  \,  \G^{abcde}\,  F_{abcde}  \ , 
\ee
where $F_{abcde}$ are  the 10d vielbein components of $F_5$   to be found. 
To extract the dilaton  factor we  assume    that the  dilaton   originates from 
integrating   out the gauge  fields $A_+,A_-$ in \rf{1}  and is  thus  given by  \ci{hms2}
(here $\M\in \psu(1,1|2)$, cf. \rf{36})
\be\la{3351}
\Phi=-\ha \STr \log \tilde\op\,,\qquad 
\qquad
\tilde\op(\M)=f \op_-(\M)f^{-1}=  f\M f^{-1}-\Omega(\M)\,.
\ee
Setting   fermions in $f$ to zero  we may split the   contributions to the $X^m$-dependent part 
of $\P$  into those of the bosonic  directions  and  the   fermionic  directions in the algebra, $e^{\Phi}= e^{\Phi_B}e^{\Phi_F}$. 
Then $\Phi_B$  is given by \rf{d1} (or its analytic continuation \rf{d2} \ci{ST}) with a particular 
$e^{\Phi_0}$ 
while $e^{\P_F}$  is proportional   to 
 $M$  or $M'$   defined in \rf{d39}   \ci{us1}
 \be \la{5533}
 e^{\P_B} = \tex -{ (1-\kappa)^2\ov 8\h} { 1 \ov \sin t \, \sin \vp} \ , \qquad \qquad 
 e^{\P_F} =  - { 4 \h \ov 1- \h^2} \, M  \ . 
 \ee
While   the \lmo for the \adst supercoset  is effectively defined in 4d target space   we shall  assume  that 
it corresponds to a $T^6$ compactification of 10d   superstring   theory, i.e. that   the corresponding   
quadratic fermionic action can be obtained  from the 10d  GS one by  embedding   both the fermions and  RR fluxes into the 10d   theory. 
Thus to extract the components of the RR 5-form we  shall assume that we can write it as in \rf{t2}, i.e.
\be\la{334} 
F_5 = \ha (1+*)F \wedge {\rm Re}\, \Omega_3 \  ,
\ee
where  $F= \ha F_{mn} dx^m \we dx^n $  and $\Omega_3$    is defined in \rf{jjj}.
The matrix $\SS^{IJ}$ in \rf{3341} can then be rewritten as
\be\la{335}
\SS^{IJ}=\tex -\epsilon^{IJ}  e^{\Phi} F_{ab}\Gamma^{ab}\Gamma^{468}\mathcal{P}_4\ ,
 \ee
where  $\mathcal{P}_4  \equiv \frac{1}{4}\left( \mathbf{1} -\Gamma^{4567}-\Gamma^{4589}-\Gamma^{6789} \right) $
is the same   projector as in \rf{aa14},\rf{a19}   (here $a,b=0,1,2,3$ and  $4, ...,9$ are   the  torus directions).
The resulting  tangent-space  components of  the RR field  $F_{ab}$ 
 that we find  from \rf{3325},\rf{3337}  are then  
\begin{small}
\be\la{336}
\begin{aligned}   
& F_{01}=\ K \,  \left(\lambda ^2-1\right)\,   \sin ^2t \sin ^2\varphi\,, \qquad \qquad \qquad   K\equiv  e^{-2\Phi_F}\ 8 i\ \cla ^{-1} \lambda ^{-8} \left(\lambda ^4-1\right)^2\ ,  \\
&F_{02}=\ha K \, \sin 2 t\,  \sinh \xi \sin \varphi \big[\left(\lambda ^2+1\right)
 \cos \varphi \cos \zeta-2 \lambda  \cos t \cosh \xi\big]\,,\\
& F_{03}=\fo   K \left(\lambda ^2-1\right) \,  \sin 2 t\, \sinh \xi \sin 2 \varphi\,  \sin \zeta\,,\\
&F_{12}= ( \l^4-1)^{-1}  K \,   \sin t\, \sin \varphi\Big[ \lambda  \left(\lambda ^4+1\right) \cos ^2\varphi \cos 2 \zeta\,  \\
&\ \ \ \ \  -  \left(\lambda ^2+1\right)^3\,  \cos t \cosh \xi \cos \varphi \cos \zeta+ \lambda \left(\left(\lambda ^4+1\right) \cos ^2t \cosh 2 \xi\, +\lambda ^4+\lambda ^2 (\cos 2 t+\cos 2 \varphi)+1\right)\Big]\,,\\
&F_{13}=\ha K \,  \sin t \sin 2 \varphi\,  \sin \zeta \big[2 \lambda  \cos \varphi \cos \zeta-\left(\lambda ^2+1\right) \cos t \cosh \xi\big]\,. 
\end{aligned}
\ee
\end{small}
Translated   to   the   algebraic coordinates and analytically continued to the ``physical"  patch \rf{patch2},\rf{yy} 
that   leads to  the same  background  as in  \rf{ff2} or \rf{ffff}  which solves   the supergravity equations.


\section{Concluding remarks}

In this paper we have found the   RR background corresponding to the \lmo for the \adst supercoset. 
We   demonstrated  that  this    background (supplemented by 6-torus directions) 
  solves the type II  supergravity equations, implying that the \lmo  
is Weyl invariant at the quantum level   and thus defines a consistent superstring theory. 

It would   be interesting  to  generalize   these results to the case of the \lmo   for $AdS_3 \times S^3$ and \adss    supercosets. 
This is  technically challenging (given the lack of isometries)   but  should be possible with some guidance 
from the   supergravity solutions   \ci{ST,ST2} that should   correspond to   the ``bosonic" version of the  $\l$-models 
associated with these higher-dimensional supercosets. 
 
 We  also  confirmed the suggestion of  \ci{us1}  that 
 there  exists a     singular  scaling limit    of the   \lmo  background that is closely related
  (classically T-dual)  to  the  analytic  continuation of the $\eta$-model. The \emo  itself  
    fails  to  be  Weyl invariant, i.e. does not correspond to a standard supergravity solution  \ci{us3}. 
It thus appears that    the 
\lmo is more  general  and better   defined than the \emo   at the quantum level. 
One   reason   is that   the \lmo  has a natural  ``first-order"   form, i.e.  is   naturally  defined  
 on a bigger space including the  gauge fields $A_\pm$  where  the   Weyl invariance should be
  manifest   (with no need   for an extra dilaton). 
A   similar    ``uplifting"    may  be   possible for the   \emo  (cf. \ci{us1}) 
and the resulting model should be Weyl invariant too 
(reflecting the fact  that the classical T-dual of the \emo,  which is also a limit of the \lmo,
  represents  a   consistent supergravity solution \ci{us3}). 

One open    question is  about the possible  interpretation of the \lmo background \rf{312}, \rf{d5}, \rf{m6}. 
The  (analytic continuation of) 4d  metric \rf{312}
 interpolates between the metric  of  $SO(1,2)/SO(2) \times SO(3)/SO(2)$  gWZW  model (for $\h=1$)   and    that of a symmetric space $AdS_2 \times H^2$ 
(for $\h=0$). 
There  are   curvature singularities on the lines\foot{Some comments on the global structure of the first 2d part of the metric \rf{312}   appeared  in \ci{ST}. Like in the gWZW  context \ci{bars}   here it  should be  more appropriate  to analyse 
the geometric structure   in terms of the  original group  variables $f$   rather than local coordinates.}
$ \h x^2 - \hi y^2=1$  and  $  \h p^2 + \hi q^2=1$.  
 Restricted to these curves  the 
 ``fermionic" factor $M$ in the dilaton\rf{d5}    is  equal to 
  $-\h - (1-\h^2) (x^2+ p^2) + 2 \sqrt{1-\h^2} xp$, i.e. the dilaton is also singular if $M\not=0$. The RR field strength  \rf{ff2} or \rf{ffff} is singular   only when $M=0$.\foot{That means the singularities of the metric and dilaton  terms  in  the l.h.s. of the Einstein equation \rf{m1} 
   cancel each other.} 
 The   metric  \rf{312}
  has no isometries (for $\h\not=0,1$) 
   but the corresponding geodesics  should be integrable (as the underlying sigma model is integrable). 
The  existence of  hidden   conserved charges should   aid the construction of physical observables  corresponding to this  geometry. 
Moreover,  like in the gWZW case,  the singularity   of the metric  seen by point-like   probes   may not be   visible in 
correlation functions for vertex operators constructed in terms of  fields in the original  \lmo action in \rf{1}.

\iffa
minor possible   issue for the future:
sentence in abstract 
The supergravity solution we find is different from the one in
arXiv:1410.1886 which should correspond to a version of the lambda-model where only the bosonic subgroup of F is gauged.

if there is no fermionic gauge symmetry  how then such model  have kappa symmetry?  But  plausible answer may be that  here  kappa symmetry will not be related   to   construction but will be an emergent symmetry  if the whole model makes sense. 
May be this is not something to worry about at this stage  but 
should be kept in mind for the future  clarification
\fi

\section*{{Acknowledgments}}
We are grateful to B. Hoare    for  very  useful  discussions  and important suggestions.
We  also   thank O. Lunin and R. Roiban  for  helpful comments. 
This work was supported by the ERC Advanced grant No.290456. 
The work of AAT   was  also supported by the 
STFC Consolidated grant  ST/L00044X/1 and RNF grant 14-42-00047.

\appendix

\section{Type II  supergravity equations in terms of RR  bispinor \\  and  \lmo  background as   10d     solution  }  \label{apa}
\def\theequation{A.\arabic{equation}}
\setcounter{equation}{0}

Here we shall rederive the  embedding   of the  \lmo background \rf{312},\rf{d5},\rf{m6} 
 into the space   of   solutions of type IIB  10d supergravity.
 We shall   use  the bispinor   notation for the  RR field strengths.

Let us  first  present
 the type II supergravity equations of motion in spinor notation starting from the superspace constraints as given in \cite{Wulff:2013kga}.
Let us consider first  the type IIA case. From eqs. (C.20) and (C.18) of \cite{Wulff:2013kga} we find (here $a,b, ...=0,1,...,9$  are  10d tangent space indices)
\begin{equation}\la{a1}\tex 
\Gamma^bU_{ab} 
+2i\nabla_\alpha\nabla_a\chi\big|_{_{\Theta=0}}
-\frac{i}{4}H_{abc}\Gamma^{bc}\Gamma_{11}\nabla_\alpha\chi\big|_{_{\Theta=0}}=0\ , 
\end{equation}
where  $\chi$ is the dilatino superfield, $H_{abc}$ is the NSNS three-form field strength   and 
\begin{equation}\la{a2}
U_{ab}=\tex \frac14\nabla_{[a}G_{b]}+\frac{1}{32}G_{[a}G_{b]}-\frac14R_{ab}{}^{cd}\Gamma_{cd}\,,\qquad
G_a=H_{abc}\Gamma^{bc}\Gamma_{11}+\mathcal S\Gamma_a\,
\end{equation}
 contains also the curvature tensor and  the   RR bispinor $\mathcal S$
 which in the IIA  case is given by ${\mathcal S} =e^\P (  \ha  \G^{ab} \G_{11}  F_{ab}  + {1 \ov 4!}  \G^{abcd} F_{abcd})$. 
 Using the expression for the torsion and for the spinor derivative of the dilatino eq.\rf{a1} becomes
\begin{equation}\la{a3}\tex 
\Gamma^bU_{ab}
+2i\nabla_aT
-\frac{i}{4}TG_a
-\frac{i}{4}H_{abc}\Gamma^{bc}\Gamma_{11}T
=0\,,
\end{equation}
where $T$ contains also the  derivative of the dilaton 
\begin{equation}\la{a4}
\tex T=\frac{i}{2}\nabla_a \Phi\,\, \Gamma^a+\frac{i}{24}H_{abc}\Gamma^{abc}\Gamma_{11}+\frac{i}{16}\Gamma^a\mathcal S\Gamma_a\,.
\end{equation}
The matrices $U_{ab}$ and $T$ appear in the integrability condition for the Killing spinor equation and in the dilatino equation respectively.
 Combining \rf{a3}   with   the same equation multiplied from left and right by $\Gamma_{11}$ gives
\begin{equation}\tex \la{a5} 
\Gamma^b\nabla_b\mathcal S\Gamma_a
-\nabla_b\Phi\,\Gamma^b\mathcal S\Gamma_a
-\frac{1}{8}H_{bcd}\Gamma^b\mathcal S\Gamma_a\Gamma^{cd}\Gamma_{11}
-\frac{1}{2}H_{abd}\Gamma^d\Gamma_{11}\mathcal S\Gamma_b
+\frac{1}{24}H_{bcd}\Gamma^{bcd}\Gamma_{11}\mathcal S\Gamma_a
=0\,.
\end{equation}
Multiplying this   by  $\Gamma^a$ from the right gives the RR sector equations of motion and the 
Bianchi identities in the form
\begin{equation}\la{a6} \tex
\Gamma^b\nabla_b\mathcal S
-\nabla_b\Phi\,\Gamma^b\mathcal S
+\frac{1}{8}H_{abc}\Gamma^a\Gamma_{11}\mathcal S\Gamma^{bc}
+\frac{1}{24}H_{abc}\Gamma^{abc}\Gamma_{11}\mathcal S
=0\,.
%
\end{equation}
The remaining equations give the Einstein equations and NSNS three-form equation of motion and Bianchi identity
\be
&& R_{ab}{}^{bc}\Gamma_c
+2\nabla_a\nabla_b\Phi\,\Gamma^b   \no \\ 
&& \tex \qquad \quad +\frac12\nabla^bH_{abc}\Gamma^c\Gamma_{11}
-\nabla^b\Phi H_{abc}\Gamma^c\Gamma_{11}
-\frac14H_{abc}H^{bcd}\Gamma_d
-\frac{1}{32}\Gamma^b\mathcal S\Gamma_a\mathcal S\Gamma_b
=0\,.
\label{a7}
\ee
The dilaton equation arises from eq. (C.19) of \cite{Wulff:2013kga} upon using \rf{a6},\rf{a7} 
\begin{align}\la{a8}
0=& \tex 
-i\Gamma^a\nabla_aT
+\frac{i}{8}\Gamma^aTG_a
+2i\nabla_a\Phi\,\Gamma^aT
-\frac{i}{24}H_{abc}\Gamma^{abc}\Gamma_{11}T
-\frac{i}{4}\mathcal ST
\nonumber\\
=&\tex
\nabla^a\nabla_a\Phi
-\nabla^a\Phi\,\nabla_a\Phi
-\frac14R_{ab}{}^{ab}
-\frac{1}{48}H_{abc}H^{abc}\,.
\end{align}
The type IIB   supergravity equations 
 take the same form but with the $32\times32$  matrices $\Gamma_a$ projected down to $16\times16$ blocks using $\frac12(1\pm\Gamma_{11})$ and
 $\Gamma_{11}$  replaced by  $\sigma_3\times I$   where $\s_3$  acts on the $SO(2)$ indices  $I,J=1,2$   of the  two MW spinors. 
The RR bispinor here   takes the form ($i\s_{2\, {IJ}} = \varepsilon_{IJ}$; see \cite{Wulff:2013kga} for further details)
  \be \la{aa1} 
\tex    \mathcal S= -\left(i \s_2 \G^a F_a   + {1 \ov 3!} \s_1  \G^{abc} F_{abc}  + {1\ov 2 \cdot 5!} i \s_2 \G^{abcde} F_{abcde}\right)\frac12(1-\Gamma_{11})
 \ . \ee

 \def \fh {H}


Let us now consider the 10d   metric corresponding to  \rf{312} (cf. \rf{t1}; here we set $T=1$ for notational simplicity) 
\be
\la{a9} &&ds^2=\frac{-dx^2+dy^2}{\fh_1(x,y)}+\frac{dp^2+dq^2}{\fh_2(p,q)}+dz^id\bar z^i \ , \\
\la{a10} &&   \fh_1=1-\h x^2+\hi  y^2\,,\qquad\qquad  \fh_2=1-\h p^2-\hi q^2\,.
\ee
The corresponding  spin connection and curvature  in terms of the   zweibein  1-forms  read
\be
&&\omega^{01}=\frac{y}{\h\sqrt {\fh_1}}e^0 -\frac{\h x}{\sqrt {\fh_1}}e^1 \,,\qquad\qquad 
\omega^{23}=-\frac{q}{\h\sqrt {\fh_2}}e^2 +\frac{\h  p}{\sqrt {\fh_2}}e^3  
\ , \\
&&R^{01}=\frac{1}{\fh_1}\big[\h+\hi-(1-\h^2)x^2+(1-\h^{-2})y^2\big] e^0\we  e^1 
\,,\\
&&R^{23}=-\frac{1 }{\fh_2}\big[ \h +\hi-(1-\h^2)p^2-(1-  \h^{-2}   )q^2\big] e^2\we e^3\,. \la{a11} 
\ee
The  metric \rf{a9}    should  be supplemented by  the dilaton in \rf{d5} 
(we  again consider the ``real" patch \rf{patch2}  where $\fh_1(x,y) <1$) 
\begin{equation}\la{a12} 
e^{2\Phi}=- e^{2\P_0} \frac{M^2}{\fh_1\fh_2}\,,\qquad \qquad 
M=\h-x^2+y^2-p^2-q^2+2\sqrt{1-\h^2}\, xp\,,
\end{equation}
which solves \rf{a8} and \rf{m3}. 
Our aim is   then to show that  the  type IIB supergravity equations are solved provided 
the above metric and dilaton are   supplemented by the 
 RR five-form field strength   given by (with all other fields  being zero) 
\be
F_5
&=&
\fo  c\,  (1+*)
\Big(
\frac{2dx\we dy+\sqrt{1-\h^2}dy\we dp}{M}
-\frac{\partial_yMdx+\partial_xMdy}{2M^2}\we dM
\Big)\we 
\mathrm{Re}\, \Omega_3
\nonumber\\
&=&
\ha  c\,   (1+*) {M^{-2}}
\Big[
\h \fh_2dx\we dy
+y(p-\sqrt{1-\h^2}x)dx\we dp
+yq dx\we dq \la{a13}
\\
&&
+\big[  \ha \sqrt{1-\h^2}(\h+x^2+y^2+p^2-q^2)-xp\big]dy\we dp
+q(\sqrt{1-\h^2}p-x)dy\we dq
\Big]\we 
\mathrm{Re}\, \Omega_3\ . \nonumber
\ee
Here $\Omega_3$ is the holomorphic three-form on $T^6$   defined in \rf{jjj}.
This  expression for $F_5$  is the same as \rf{t2}   with $F_{mn}$ given by \rf{ff2}.

The   corresponding  RR bispinor \rf{aa1} 
 is then given by 
  (here  projection from the right by $\frac12(1-\Gamma_{11})$  as in \rf{aa1} is understood)
\begin{align}\la{a14}
\mathcal S=& i\sigma^2 \ \td { \mathcal S} \ , \qquad \qquad 
\td{ \mathcal S}= -{\tex \frac{1}{2\cdot5!}} e^\Phi\,  F_{abcde}\Gamma^{abcde} \ , 
\\
\td { \mathcal S} =&
  c e^{\P_0}\,
\Big(
\sqrt{1-\h^2}\, \Gamma^{12}
+2\sqrt{-{\fh_1}/{\fh_2}}\, \Gamma^{01}
-\frac{\partial_1M\partial_aM\Gamma^{0a}+\partial_0M\partial_aM\Gamma^{1a}}{2M\sqrt{-\fh_1\fh_2}}
\Big)
\Gamma^{468}\mathcal P_4\,. \la{aa14}
%
\end{align}
We have defined the projector
\begin{equation}\tex 
\mathcal P_4=\frac14(1-i\slashed J\Gamma^{(7)})=\frac14(1-\Gamma^{4567}-\Gamma^{4589}-\Gamma^{6789})\,.
\la{a19}
\end{equation}
Here 
 $\slashed J=\Gamma^{45}+\Gamma^{67}+\Gamma^{89}$ is the K\"ahler form on $T^6$ contracted with gamma matrices and $\Gamma^{(7)}=i\Gamma^{456789}$. The fact that $(i\slashed J\Gamma^{(7)})^2+2i\slashed J\Gamma^{(7)}-3=0$ means that $i\slashed J\Gamma^{(7)}\frac12(1\pm\Gamma_{11})$ has 12 eigenvalues equal to $1$ and four equal to $-3$. Note also that $\mathcal P_4\Gamma_{a'}\mathcal P_4=0$ where $a'=4,\ldots,9$.

One can then check that 
the supergravity equations and Bianchi identities for the RR fields (\ref{a6}) are satisfied, namely, 
\begin{equation}\la{a15} \tex 
\Gamma^a\nabla_a\mathcal S-\nabla_a\Phi\,\Gamma^a\mathcal S
=
\Gamma^a\partial_a\mathcal S
-\frac14\Gamma^a\omega_a{}^{bc}[\Gamma_{bc},\mathcal S]
-\partial_a\Phi\,\Gamma^a\mathcal S
=
0\,.
\end{equation}
%
The Einstein equations \rf{a7} for $H_{abc}=0$ simplify to (projection  from the right by $\frac12(1+\Gamma_{11})$ is suppressed)
\begin{equation}\la{a21} 
\tex 
R_{ab}{}^{bc}\Gamma_c+2\nabla_a\nabla_b\Phi\,\Gamma^b-\frac{1}{32}\Gamma^b\mathcal S\Gamma_a\mathcal S\Gamma_b=0\ . 
%
\end{equation}
These are  also  satisfied
provided 
\be \la{a16}  c^2= 16 \h^{-1} e^{-2\P_0}   \ ,  \ee 
which is in agreement with \rf{m6},\rf{ff2} (in this Appendix  $T=1$).


\section{Realisation of $\su(1,1|2)$ superalgebra}\label{apb}
\def\theequation{B.\arabic{equation}}
\setcounter{equation}{0}


The superalgebra $\hat{\alg{f}}=\su(1,1|2)$  is represented by 
$4\times 4$ matrices $\M$ which satisfy $\STr \M=0$ and the reality condition $\M^\dagger H+H\M=0$, with $H=\text{diag}(\sigma_3,\mathbf{1}_2)$. 
The $\mathbb{Z}_4$ automorphism $\Upsilon$ 
\be\la{b1}
\Upsilon(\M)=
-\left(\begin{array}{cc}\sigma_3 &\mathbf{0} \\ \mathbf{0} & \sigma_3\end{array}\right)
\left(\begin{array}{cc}m_{11}^t &-m_{21}^t \\ m_{12}^t & m_{22}^t\end{array}\right)
\left(\begin{array}{cc}\sigma_3 &\mathbf{0} \\ \mathbf{0} & \sigma_3\end{array}\right)\,,
\qquad
\M=\left(\begin{array}{cc}m_{11} &m_{12} \\ m_{21} & m_{22}\end{array}\right)\,,
\ee
identifies four subspaces $\hat{\alg{f}}^{(k)}$ labeled by $k=0,\ldots,3$ depending on the eigenvalue of $\Upsilon$ on an element $\M\in \hat{\alg{f}}^{(k)},\ \Upsilon(\M)=i^k \M$.
We define the supertrace as $\STr(\M)=\sum_{j=1}^2 \M_{jj}-\sum_{j=3}^4 \M_{jj}$.

We will  
 realise the $\su(1,1|2)$ algebra in terms of explicit $4\times 4$ matrices.
In the upper-left $2\times 2$ block we place generators of AdS$_2$, while we put generators of S$^2$ in the lower-right one. The off-diagonal blocks contain the odd generators of the algebra. We denote by $\gen{P}_a, \gen{J}_{ab}$ the bosonic generators of the algebra, where indices $a=0,1$ are used for AdS$_2$ and $a,b=2,3$ for S$ ^2$. We define them as
\be\la{3.1}
\begin{aligned}
&\gen{P}_{{a}} = 
\left(
\begin{array}{cc}
 -\frac{1}{2} \check{\gamma}_{{a}} & \mathbf{0}  \\
 \mathbf{0} & \mathbf{0} \\
\end{array}
\right), \quad {a}=0,1,
\qquad
&&\gen{P}_{ a} = 
\left(
\begin{array}{cc}
 \mathbf{0} & \mathbf{0}  \\
 \mathbf{0} & \frac{i}{2} \hat{\gamma}_{{a}} \\
\end{array}
\right), \quad {a}=2,3 \ , 
\\
&\gen{J}_{ab} = 
\left(
\begin{array}{cc}
 \frac{1}{2} \check{\gamma}_{ab} & \mathbf{0}  \\
 \mathbf{0} & \mathbf{0} \\
\end{array}
\right), \quad a,b=0,1,
\qquad
&&\gen{J}_{ab} = 
\left(
\begin{array}{cc}
 \mathbf{0} & \mathbf{0}  \\
 \mathbf{0} & \frac{1}{2} \hat{\gamma}_{ab} \\
\end{array}
\right), \quad a,b=2,3,
\end{aligned}
\ee
where
\be
\la{3.2} 
\tex 
\check{\gamma}_0= i \sigma_3\,, \quad \check{\gamma}_1= \sigma_2\,,
\quad \hat{\gamma}_{2}= - \sigma_3\,,\quad \hat{\gamma}_{3}= - \sigma_2\,,
\qquad 
\check{\gamma}_{ab} = \frac{1}{2} [\check{\gamma}_a,\check{\gamma}_b]\,,
\quad \hat{\gamma}_{ab} = \frac{1}{2} [\hat{\gamma}_a,\hat{\gamma}_b]\ . 
\ee
To define   the odd generators we use the matrices $\gen{Q}^{I\check \alpha \hat \alpha}$, where $I=1,2$ 
and  $\check \alpha ,\hat \alpha=1,2$ are  spinor indices in AdS$_2$ and S$^2$ respectively 
\be\label{3.3}
\begin{aligned}
\gen{Q}^{I\check \alpha \hat \alpha} &=\frac{e^{+i\pi/4}}{\sqrt{2}}
\left(
\begin{array}{cc}
 \mathbf{0} & m^{I\check \alpha \hat \alpha}  \\
 -\sigma_3 \left(m^{I\check \alpha \hat \alpha}\right)^\dagger \sigma_3 & \mathbf{0} \\
\end{array}
\right),
\\
m^{1\check \alpha \hat \alpha} &= -e^{+i\pi/4} \, (-1)^{\hat\alpha}\, u^{\check \alpha\hat \alpha} ,
\qquad\qquad
m^{2\check \alpha \hat \alpha} = e^{-i\pi/4} \,  (-1)^{\hat\alpha}\, u^{\check \alpha\hat \alpha} ,
\end{aligned}
\ee
where $u^{\check \alpha\hat \alpha}$ is the $2\times 2$ 
 matrix with zero everywhere except the element 1 at position $(\check \alpha,\hat \alpha)$.
Considering the Grassmann envelope of the superalgebra and demanding that $\gen{Q}^{I\check\alpha\hat\alpha}\theta_{I\check\alpha\hat\alpha}$ satisfies the reality condition of $\su(1,1|2)$ we find that the fermions $\theta_I$ satisfy the Majorana condition\footnote{We will be omitting spinor indices most of the time. When we need to reintroduce them we assume that gamma matrices for AdS$_2$ $(\check\gamma_a)_{\check\alpha}^{\ \check\beta}$ are acting only on checked indices of the fermions $\theta_{I\check\alpha\hat\alpha}$, while gamma matrices for S$^2$ $(\hat\gamma_a)_{\hat\alpha}^{\ \hat\beta}$ are acting only on their hatted indices.}
\be\la{3.44}
\bar{\theta}_I = \theta_I^\dagger (\check\g^0\otimes\mathbf{1}_2) =  \theta_I^t (\sigma_3 \otimes \sigma_3).
\ee
Then $\gen{J}_{ab}$ and $\gen{P}_a$ belong to the subspaces of grading 0 and 2 respectively, while $\gen{Q}^{1\check \alpha \hat \alpha}$ and $\gen{Q}^{2\check \alpha \hat \alpha}$ to the ones of grading 1 and 3.
The commutation relations can be read off by computing explicitly the matrix multiplications 
\be
&& \text{AdS}_2 : \quad [ {\gen{P}}_a, {\gen{P}}_b ] = {\gen{J}}_{ab}, \qquad [ {\gen{P}}_{a}, {\gen{J}}_{bc} ] = \eta_{a[b} {\gen{P}}_{c]}\ , \\
&& \text{S}^2 : \quad [ {\gen{P}}_a, {\gen{P}}_b ] = -{\gen{J}}_{ab}, \qquad  [ {\gen{P}}_{a}, {\gen{J}}_{bc} ] = \eta_{a[b} {\gen{P}}_{c]} \ ,\\
&&\tex  [\gen{Q}^{I} \theta_I, \gen{P}_a] = - \frac{i}{2} \epsilon^{IJ} \gen{Q}^{J} \bg_a  \theta_I,  \qquad
 [\gen{Q}^{I} \theta_I, \gen{J}_{ab}] =  -\frac{1}{2} \delta^{IJ} \gen{Q}^{J} \bg_{ab}  \theta_I, \\
 &&\tex \la{eq:comm-rel-QQ} [ \gen{Q}^{I} \lambda_I, \gen{Q}^{J} \psi_J ] = \, i \, \delta^{IJ} \bar{\lambda}_I \bg^a  \psi_J \ \gen{P}_a  %
-   \epsilon^{IJ} \bar{\lambda}_I (\bg^{01} {\gen{J}}_{01} -\bg^{23}  {\gen{J}}_{23}) \psi_J \ 
 - \frac{i}{2} \delta^{IJ} \bar{\lambda}_I \psi_J \mathbf{1}\ ,  \la{3.4}
\ee
where it was convenient to introduce the $4\times 4$ matrices\footnote{These matrices $\bg_a$ are not gamma matrices since they do not satisfy the Clifford algebra relations when we mix indices from AdS$_2$ and S$^2$. However,  they appear naturally in the supercoset construction and they have a natural embedding in the $32\times 32$ gamma matrices, see Appendix~\ref{apd}.}
\be\label{3.5}
\begin{aligned}
& \bg_a = \check{\g}_a \otimes \mathbf{1}_2,
\quad a=0,1,
\qquad\qquad \qquad 
 \bg_a =  \mathbf{1}_2 \otimes i\hat{\g}_a,
\quad a=2,3, \\
& \bg_{ab} = \check{\g}_{ab} \otimes \mathbf{1}_2,
\quad a,b=0,1,
\qquad\ \ \qquad
 \bg_{ab} =  \mathbf{1}_2 \otimes \hat{\g}_{ab},
\quad a,b=2,3. \\
\end{aligned}
\ee
To get $\psu(1,1|2)$ from $\su(1,1|2)$ one simply needs to project out the generator proportional to the identity $\mathbf{1}$.


In the  above  basis for $\su(1,1|2)$ 
 we find the following bilinear form induced by the supertrace
\be\la{b2}
\begin{aligned}
&\Str[{\gen{J}}_{01}{\gen{J}}_{01}]=  \ha \ ,  \qquad
&&\Str[\gen{P}_a\gen{P}_b]=\ha  \eta_{ab}\ , \\
&\Str[{\gen{J}}_{23}{\gen{J}}_{23}]= \ha\ ,  \qquad
&&\Str[\gen{Q}^I \lambda_I \, \gen{Q}^J \psi_J ]= - \epsilon^{IJ} \bar{\lambda}_I \psi_J = - \epsilon^{JI} \bar{\psi}_J \lambda_I \,.
\end{aligned}
\ee
We define the matrix 
\be\label{b3}
W=\text{diag}(1,1,-1,-1)\,,
\ee
which is not an element of $\psu(1,1|2)$, but plays an important role in the computation of the kappa-symmetry variation of the world-sheet metric  in \rf{310}.

With a group element of the form~\eqref{3200}, we find that the Maurer-Cartan form is
\be\label{b4}
\begin{aligned}
\CF^{-1}d\CF=&\tex   \left( e^a -\frac{i}{2}  \bar{\theta}_I  \bg^a  D^{IJ} \theta_J \right) \gen{P}_a 
 + \gen{Q}^{I} \, D^{IJ} \theta_J \\
&\tex  -\frac{1}{2} \omega^{ab} {\gen{J}}_{ab} + \frac{1}{4} \epsilon^{IJ} \bar{\theta}_I \left( \bg^{01} {\gen{J}}_{01} - \bg^{23} {\gen{J}}_{23}  \right)  D^{JK} \theta_K 
+\mathcal{O}(\theta^3)\,.
\end{aligned}
\ee
Here the operator $D^{IJ}$  defined on fermions $\theta$ is
\be\label{b5}\tex 
D^{IJ} = \delta^{IJ} \left( d - \frac{1}{4} \omega^{ab} \bg_{ab}  \right)
- \frac{i}{2} \epsilon^{IJ} e^a  \bg_a\  ,
\ee
where  $e^a, \omega^{ab}$ are the vielbein and the spin-connection of the undeformed AdS$_2\times$S$^2$ supercoset.
The explicit  form of  the Maurer-Cartan form is necessary to derive most of the ingredients  needed 
to construct the $\lambda$-model, from the solution of the gauge fields $A_\pm$ to the $\k$-symmetry transformations upon the formal substitution of the derivative $d$ with the variation $\delta$.

\section{Relations for $\op_\pm$   operators in $\k$-symmetry variations}\label{apc}
\def\theequation{C.\arabic{equation}}
\setcounter{equation}{0}
Here 
 we collect some results on the linear operators $\op_\pm$ defined in~\eqref{36} which are 
 needed for the explicit calculations in section \ref{next}.
To start, we expand $\op_\pm$ and its  inverse  $\op_\pm^{-1}$ in powers of fermions as
\be\la{c1}
\begin{aligned}
\op_\pm &= \op^{(0)}_\pm + \op^{(1)}_\pm+ \op^{(2)}_\pm + \ldots\ , 
\qquad \qquad 
\op^{-1}_\pm&= \op^{inv,(0)}_\pm +\op^{inv,(1)}_\pm +\op^{inv,(2)}_\pm +\ldots\ , 
\end{aligned}
\ee
where one has the obvious relations\foot{For  the computation of the   $\k$-symmetry variations
in section  3  it will be enough to stop at linear order in fermions
but to determine  the quadratic fermionic action would require to go to quadratic order.}
\be\la{c2} 
\op^{inv,(0)}_\pm =(\op^{(0)}_\pm)^{-1}\,,
\qquad\qquad 
\op^{inv,(1)}_\pm = - (\op^{(0)}_\pm)^{-1} \circ \op^{(1)}_\pm \circ (\op^{(0)}_\pm)^{-1}, \ \ ...
\ee
Let us define  the matrices $(k_\pm)_i^{\ j}$ as
\be\la{c3}
\op^{(0)}_\pm(\gen{T}_i) = (k_\pm)_i^{\ j} \gen{T}_j\,,
\ee
where $\gen{T}_i$ denotes any  (bosonic  or  fermionic) generator of the superalgebra.
 It is easy to see that the matrices $k_\pm$  can be put into  block form    
 where each  of the three blocks mixes  only the generators of AdS$_2$, or of S$^2$, or odd generators,  respectively and 
all  blocks are  invertible. Then 
\be\la{c4}
(\op^{(0)}_\pm)^{-1}(\gen{T}_i) = (k_\pm^{-1})_i^{\ j} \gen{T}_j\,, \qquad \qquad (k_\pm^{-1})_i^{\ j}(k_\pm)_j^{\ k}=\delta_i^{\ k}\ . 
\ee
\iffa 
{\bf 
For the action of the operators at higher powers of fermions we use a different letter to denote the matrix and we write
\be
\op^{(j)}_\pm(\gen{T}_a) = (\ell_\pm^{(j)})_a^{\ b} \gen{T}_b\,,
\qquad
j=1,2,\ldots
\ee
So we get for example
\be
\op^{inv,(1)}_\pm(\gen{T}_a) = -(k^{-1}_\pm)_a^{\ b} (\ell^{(1)}_\pm)_b^{\ c} (k^{-1}_\pm)_c^{\ d} \gen{T}_d\,.
\ee
}
\fi
Let us present the expression   for $k_\pm $  written  in block form 
\be k_{\pm,A}\oplus k_{\pm,S}\oplus k_{\pm,F}
\ee 
in algebraic coordinates  where they  take a more compact form.\footnote{The results of this appendix have been simplified by assuming $y>0,q>0$.} 
In the basis $\{\gen{P}_0,\gen{P}_1,\gen{J}_{01}\}$ for the generators of AdS$_2$ we find
\be
 k_{\pm,A}=\left(
\begin{array}{ccc}
 \frac{2 \kappa }{\kappa -1}+\frac{2 (\kappa +1) y^2}{\kappa  (\pm 1+\kappa )} & -\frac{2 x y (\kappa
   +1)}{\pm 1+\kappa } & -\frac{2 y (\kappa +1) \sqrt{H_1}}{\sqrt{\kappa } (\pm 1+\kappa )} \\
 -\frac{2 x y (\kappa +1)}{\pm 1+\kappa } & \frac{2 \kappa }{\kappa -1}+\frac{2 (\kappa +1) \left(y^2-\kappa 
   H_1\right)}{(\pm 1+\kappa ) \kappa } & \frac{2 x \sqrt{\kappa } (\kappa +1) \sqrt{H_1}}{\pm 1+\kappa } \\
\pm  \frac{2  y \sqrt{H_1}}{\sqrt{\kappa }} &  \mp 2 x \sqrt{\kappa } \sqrt{H_1} & \mp 2  H_1 \\
\end{array}
\right), \quad    H_1=1-\kappa  x^2+\kappa^{-1} y^2  \quad 
\ee
In the basis $\{\gen{P}_2,\gen{P}_3,\gen{J}_{23}\}$ for the generators of S$^2$ we find
\be
k_{\pm,S}=\left(
\begin{array}{ccc}
 \frac{2 \kappa }{\kappa -1}-\frac{2 q^2 (\kappa +1)}{\kappa  (\pm 1+\kappa )} & -\frac{2 p q (\kappa
   +1)}{\pm 1+\kappa } & \frac{2 q (\kappa +1) \sqrt{H_2}}{\sqrt{\kappa } (\pm 1+\kappa )} \\
 \frac{2 p q (\kappa +1)}{\pm 1+\kappa } & \frac{2 \kappa }{\kappa -1}-\frac{2 (\kappa +1) \left(q^2+\kappa 
   H_2\right)}{\kappa  (\pm 1+\kappa )} & -\frac{2 p \sqrt{\kappa } (\kappa +1) \sqrt{H_2}}{\pm 1+\kappa } \\
\pm  \frac{2  q \sqrt{H_2}}{\sqrt{\kappa }} & \pm 2 p  \sqrt{\kappa } \sqrt{H_2} & \mp 2  H_2 \\
\end{array}
\right)  ,  \quad  H_2=1-\kappa  p^2-\kappa^{-1} q^2  \quad 
\ee
Let us   order the odd generators in \rf{3.3}  as
$\{\gen{Q}^{111},\gen{Q}^{112},\gen{Q}^{121},\gen{Q}^{122},$
$\gen{Q}^{211},\gen{Q}^{212},\gen{Q}^{221},\gen{Q}^{222}\}$ 
 and  decompose the matrices\footnote{It is assumed that their action with explicit indices is as $\op_\pm^{(0)}(\gen{Q}^{I\check\alpha\hat\alpha}) = (k_{\pm,F})^{I\check\alpha\hat\alpha}_{\ J\check\beta\hat\beta}\gen{Q}^{J\check\beta\hat\beta}$.} $k_{\pm,F}$ as
\be\label{c8}
k_{\pm,F} = \sum_{\mu=0}^3\sum_{\check a=0}^3\sum_{\hat a=0}^3 \ c^F_\pm(\mu,\check a,\hat a) \ \ \mathbf{s}_\mu \otimes \mathbf{g}_{\check a} \otimes \mathbf{g}_{\hat a}\,,
\ee
where we have defined 
\be
\begin{aligned}
\mathbf{s}_\mu &= \{ \mathbf{1}_2, \sigma_1, i \sigma_2, \sigma_3 \}\,, \quad
\mathbf{g}_{\check a} &= \{ \mathbf{1}_2, \check\gamma_0^t, \check\gamma_1^t,-\check\gamma_{01}^t \}\,, \quad
\mathbf{g}_{\hat a} &= \{ \mathbf{1}_2, \hat\gamma_2^t, \hat\gamma_3^t,-\hat\gamma_{23}^t\}\,.
\end{aligned}
\ee
The coefficients $c^F_\pm$ are different for the two operators and are given by
\begin{small}
\be\nonumber
\begin{aligned}
&\tex c^F_+(0,0,0)= \kappa  p x-\frac{1}{\sqrt{1-\kappa ^2}},\
&&c^F_+(0,0,3)= -q x, \
&&&c^F_+(0,1,1)= i   \sqrt{H_1} \sqrt{H_2}, \\
&c^F_+(0,3,0)= p y, 
&&c^F_+(0,3,3)= -\frac{q y}{\kappa },
&&&\tex c^F_+(2,0,1)= -\sqrt{\kappa   } x \sqrt{H_2}, \\
&c^F_+(2,1,0)= i \sqrt{\kappa } p \sqrt{H_1},
&&\tex c^F_+(2,1,3)= -\frac{i q   \sqrt{H_1}}{\sqrt{\kappa }}, 
&&&\tex c^F_+(2,3,1)= -\frac{y \sqrt{H_2}}{\sqrt{\kappa }}, \\
&\tex c^F_+(3,0,0)= \frac{\kappa   }{\sqrt{1-\kappa ^2}},
\\
&\tex c^F_-(0,0,0)= 1-\frac{\kappa  p x}{\sqrt{1-\kappa ^2}},\qquad
&&\tex c^F_-(0,0,3)= \frac{q x}{\sqrt{1-\kappa   ^2}},
&&&\tex c^F_-(0,1,1)= -i \frac{\sqrt{H_1} \sqrt{H_2}}{\sqrt{1-\kappa ^2}} , \\
&\tex c^F_-(0,3,0)= -\frac{p   y}{\sqrt{1-\kappa ^2}}, 
&&\tex c^F_-(0,3,3)= \frac{q y}{\sqrt{\kappa ^2-\kappa ^4}},
&&&\tex   c^F_-(1,0,1)= \kappa    \sqrt{\frac{\kappa }{1-\kappa ^2}} x \sqrt{H_2}, \\
&\tex  c^F_-(1,1,0)= -i \kappa  \sqrt{\frac{\kappa }{1-\kappa ^2}} p   \sqrt{H_1},
&&\tex  c^F_-(1,1,3)= i \sqrt{\frac{\kappa }{1-\kappa ^2}} q \sqrt{H_1},
&&&\tex  c^F_-(1,3,1)= \sqrt{\frac{\kappa   }{1-\kappa ^2}} y \sqrt{H_2}, \\
&\tex  c^F_-(2,0,1)= \sqrt{\frac{\kappa }{1-\kappa ^2}} x \sqrt{H_2},
&&\tex  c^F_-(2,1,0)= -i   \sqrt{\frac{\kappa }{1-\kappa ^2}} p \sqrt{H_1},
&&&\tex  c^F_-(2,1,3)= \frac{i  q   \sqrt{H_1}}{\sqrt{\kappa(1-\kappa ^2)} }, \\
&\tex  c^F_-(2,3,1)= \frac{y   \sqrt{H_2}}{\sqrt{\kappa  \left(1-\kappa ^2\right)}} ,
&&\tex   c^F_-(3,0,0)= -\frac{\kappa ^2 p x}{\sqrt{1-\kappa ^2}},
&&&\tex  c^F_-(3,0,3)= \frac{\kappa  q   x}{\sqrt{1-\kappa ^2}}, \\
&\tex  c^F_-(3,1,1)= -i \kappa  \frac{\sqrt{H_1}   \sqrt{H_2}}{\sqrt{1-\kappa ^2}} ,
&&\tex  c^F_-(3,3,0)= -\frac{\kappa  p y}{\sqrt{1-\kappa ^2}},
&&&\tex c^F_-(3,3,3)= \frac{q y}{\sqrt{1-\kappa   ^2}}.
\end{aligned}
\ee
\end{small}


Let us now   demonstrate  how one can 
  put the $\k$-symmetry variations into the standard form. 
  We need to compute~\rf{34}, where $f^{-1}\delta f$ is obtained from~\rf{b4} by formally substituting the derivative $d$ with the variation $\delta$, and by assigning weights 0 and 1 to the variations $\delta\theta_I$ and $\delta X^m$  at leading order in  expansion in  fermions.
When we project on odd generators we get equations for $\delta\theta_I$. At leading order in  fermions the l.h.s. 
 of these equations is just $(\op_+^{(0)})^{-1}(\gen{Q}\delta \theta)=\gen{Q}(k_{+,F}^{-1})^t\delta\theta$, where $t$ is transposition.
From the r.h.s.  of~\eqref{34} we find
\be
\begin{aligned}
&A_-^{(2)}\tilde\alpha+\tilde\alpha A_-^{(2)}
=-\ha \cla^{-1}  \gen{Q}^{1}({E}^{(-)\check{a}}_-\bg_{\check{a}}-{E}^{(-)\hat{a}}_-\bg_{\hat{a}} )\k_{1}\,,
\\
&A_+^{(2)}\hat\alpha+\hat\alpha A_+^{(2)}=+\ha \cla^{-1} \gen{Q}^{2}({E}^{(+)\check{a}}_+\bg_{\check{a}}-{E}^{(+)\hat{a}}_+\bg_{\hat{a}} )\k_{2}\,,
\end{aligned}
\la{kappaI}
\ee
where  $\check{a}$   are   AdS indices    and $\hat{a}$ are   sphere  ones.  We used that 
(as can be shown with our realisation of the superalgebra)
\be
\gen{Q}^I \gen{P}_{\check{a}} + \gen{P}_{\check{a}} \gen{Q}^I = -\ha  \gen{Q}^I \bg_{\check{a}}\,,
\qquad\qquad 
\gen{Q}^I \gen{P}_{\hat{a}} + \gen{P}_{\hat{a}} \gen{Q}^I = +\ha  \gen{Q}^I \bg_{\hat{a}}\,.
\ee
We first redefine $\theta \to  \cla ^{-1/2}\, k_{+,F}^t \,  \theta,$
and $ \vkappa \to 2 \cla ^{1/2}\vkappa\ $,
finding that the $\k$-symmetry transformations for the fermions become
$
\delta\theta_1
=-  ({E}^{(-)\check{a}}_-\bg_{\check{a}}-{E}^{(-)\hat{a}}_-\bg_{\hat{a}} )\k_{1}\,,
$ and $
\delta\theta_2
=  ({E}^{(+)\check{a}}_+\bg_{\check{a}}-{E}^{(+)\hat{a}}_+\bg_{\hat{a}} )\k_{2}\,.
$
These variations differ only in  the choice of the vielbein, meaning that it is enough to redefine $\theta_1\to U\theta_1$ and $\k_1\to -U\k_1$ with $U=-\sigma_2\otimes \sigma_2$ such that  $U^{-1}\bg_a U = \Lambda_a^{ \ b} \bg_b$ to obtain
\be
&\delta\theta_1
=  ({E}^{\check{a}}_-\bg_{\check{a}}-{E}^{\hat{a}}_-\bg_{\hat{a}} )\k_{1}\,,
\qquad
&\delta\theta_2
=  ({E}^{\check{a}}_+\bg_{\check{a}}-{E}^{\hat{a}}_+\bg_{\hat{a}} )\k_{2}\,,
\ee
where  $E=E^{(+)}$.
This is the desired  standard form of the $\k$ symmetry    variations  which can be rewritten  also 
in the 10d notation as in~\rf{328}.

There is still a  freedom to rescale  $\theta_I\to c_I \theta_I$ and $\k_I\to c_I\k_I$.
Then  $c_1=\lambda,\  c_2=1$  are fixed  by requiring that the $\k$-symmetry transformations for the 
bosons are  also of the  standard  form. 
These  are obtained by projecting~\eqref{34} on generators of grading 2, keeping only the leading order contributions.
After  taking into account the previous redefinitions for  $\theta$ we find\footnote{Here we choose to omit  the rather long 
explicit expressions for the coefficients $c^a_{b_1\ldots b_n}$.}
\be
&&\la{c12} 
 \qquad \qquad\qquad \qquad
 \qquad E^a_m\delta X^m = \tex \frac{i}{2} \bar{\theta}_I \bg^a \delta \theta_I + 2 \bar{\theta}_I \Delta^a_{IJ} \delta \theta_J\,,
\\
&&\Delta^a_{IJ}=\delta^{IJ}\left(c_{012}^a\bg^{012}+c_{023}^a\bg^{023}\right)
+\epsilon^{IJ} \left( c_{02}^a\bg^{02}+c_{03}^a\bg^{03}+c_{12}^a\bg^{12}+c_{13}^a\bg^{13}+c_{0123}^a\bg^{0123} \right)\,.\no 
\ee
The last term in \rf{c12} 
 can be  canceled by a shift  $X^m\to X^m + E^m_a \bar{\theta}_I \Delta^a_{IJ} \theta_J$. This is possible thanks to the symmetry property of $\Delta^a_{IJ}$ under transposition of the spinor indices and labels $I,J$, which makes this shift non-vanishing.\footnote{Notice that in general the symmetry property of $\bg_{m_1\ldots m_n}$ does not need to be the same as  of $\Gamma_{m_1\ldots m_n}$,
 see end of Appendix~\ref{apd} for definitions.}

The $\k$-symmetry transformations of the world-sheet metric are obtained from~\eqref{310}. One needs to compute $A_\pm$ by implementing the above  redefinitions of the fermions. The  redefinitions of the bosons do not modify the result at the
 leading order. Notice that the non-vanishing contribution 
 comes from the last term in the commutation relation~\eqref{eq:comm-rel-QQ}, similarly to what happens 
 in the undeformed supercoset case.

\section{Gamma matrices}\label{apd}
\def\theequation{D.\arabic{equation}}
\setcounter{equation}{0}

The basis for the $32 \times 32$ gamma matrices  that we use  is
\be\la{ad1}
\begin{aligned}
&\Gamma_0 = \sigma_1 \otimes \sigma_3 \otimes \mathbf{1}_2 \otimes \mathbf{1}_2 \otimes \mathbf{1}_2 \,,
\qquad
&&\Gamma_1 = \sigma_1 \otimes \sigma_2 \otimes \mathbf{1}_2 \otimes \mathbf{1}_2 \otimes \mathbf{1}_2 \,,\\
&\Gamma_2 = \sigma_2 \otimes \mathbf{1}_2 \otimes \sigma_3 \otimes \mathbf{1}_2 \otimes \mathbf{1}_2 \,,
\qquad
&&\Gamma_3 = \sigma_2 \otimes \mathbf{1}_2 \otimes \sigma_2 \otimes \mathbf{1}_2 \otimes \mathbf{1}_2 \,,\\
&\Gamma_4 = \sigma_2 \otimes \mathbf{1}_2 \otimes \sigma_1 \otimes \mathbf{1}_2 \otimes \sigma_1 \,,
\qquad
&&\Gamma_5 ={\phantom{-}} \sigma_1 \otimes \sigma_1 \otimes \mathbf{1}_2 \otimes \sigma_1 \otimes \mathbf{1}_2 \,,
\\
&\Gamma_6 = \sigma_2 \otimes \mathbf{1}_2 \otimes \sigma_1 \otimes \mathbf{1}_2 \otimes \sigma_2 \,,
\qquad
&&\Gamma_7 =- \sigma_1 \otimes \sigma_1 \otimes \mathbf{1}_2 \otimes \sigma_2 \otimes \mathbf{1}_2 \,,
\\
&\Gamma_8 = \sigma_2 \otimes \mathbf{1}_2 \otimes \sigma_1 \otimes \mathbf{1}_2 \otimes \sigma_3 \,,
\qquad
&&\Gamma_9 = {\phantom{-}}\sigma_1 \otimes \sigma_1 \otimes \mathbf{1}_2 \otimes \sigma_3 \otimes \mathbf{1}_2 \,.
\end{aligned}
\ee
These $\G_a$  satisfy
\be\la{ad2}
\{ \Gamma_a,\Gamma_b\} = 2 \eta_{ab} \mathbf{1}_{32}\,,
\qquad
\text{Tr}(\Gamma_a)=0\,,
\qquad
(\mathcal{C}\Gamma_m)^t = + \mathcal{C}\Gamma_m\,,
\ee
with
$\mathcal{C}=i \sigma_2 \otimes \sigma_3 \otimes \sigma_3 \otimes \sigma_2 \otimes \sigma_2$.

The gamma matrices  corresponding to  AdS$_2$ and S$^2$ are  
\be \la{ad3} 
\Gamma_a = \sigma_1 \otimes \check \gamma_a \otimes \mathbf{1}_2 \otimes \mathbf{1}_2 \otimes \mathbf{1}_2, \ \  \ a=0,1; \quad \quad
\Gamma_a = \sigma_2 \otimes \mathbf{1}_2 \otimes \hat\gamma_a \otimes \mathbf{1}_2 \otimes \mathbf{1}_2 \ , \ \ \ a=2,3  \ . \ee
With  this definition we have 
\be\la{ad4}
\bar \T_I \Gamma_a \T_J=\bar\theta_I \bg_a \theta_J\,,
\ee
where $\bg_a$ are defined in~\eqref{3.5} 
and  the 32-component spinors $\T_I$ are related to the 4-component  $\theta_I$ as in \rf{329}
\be\la{ad5} \tiny
\T_I = \left(\begin{array}{c} 1 \\ 0 \end{array}\right) \otimes \theta_I \otimes \left(\begin{array}{c} 1 \\ 0 \\ 0 \\ 0 \end{array}\right)\,.
\ee
The conjugate   fermions are  defined 
 by $\bar \T_I = \T_I^t \mathcal{C}$ and $\bar \theta_I = \theta_I^t (\sigma_3\otimes \sigma_3)$.
More generally, we define $\bg_{m_1\ldots m_n}$ with $n$ indices by requiring that $\bar \T_I \Gamma_{m_1\ldots m_n} \T_J=\bar\theta_I \bg_{m_1\ldots m_n} \theta_J$ when $n$ is odd, and $\bar K_I \Gamma_{m_1 \ldots m_n} \T_J=\bar\k_I \bg_{m_1 \ldots m_n} \theta_J$ when $n$ is even.




\begin{thebibliography}{30}
\parskip=0.2 pt


\bibitem{dmv}
F.~Delduc, M.~Magro and B.~Vicedo,
``An integrable deformation of the $AdS_5 \times S^5$ superstring action,''
Phys.\ Rev.\ Lett.\ {\bf 112}, no. 5, 051601 (2014)
[\arxivlink{1309.5850}].
``Derivation of the action and symmetries of the $q$-deformed $AdS_5 \times S^5$ superstring,''
JHEP {\bf 1410} (2014) 132
[\arxivlink{1406.6286}].


\bibitem{Klimcik:2002zj}
C.~Klimcik,
``Yang-Baxter sigma models and dS/AdS T duality,''
JHEP {\bf 0212}, 051 (2002)
[\arxivlink{hep-th/0210095}].
``On integrability of the Yang-Baxter sigma-model,''
J.\ Math.\ Phys.\ {\bf 50}, 043508 (2009)
[\arxivlink{0802.3518}].
``Integrability of the bi-Yang-Baxter sigma-model,''
Lett.\ Math.\ Phys.\ {\bf 104}, 1095 (2014)
[\arxivlink{1402.2105}].


\bibitem{hms1}
T.~J.~Hollowood, J.~L.~Miramontes and D.~M.~Schmidtt,
``Integrable Deformations of Strings on Symmetric Spaces,''
JHEP {\bf 1411} (2014) 009
[\arxivlink{1407.2840}].

\bibitem{hms2}
T.~J.~Hollowood, J.~L.~Miramontes and D.~M.~Schmidtt,
``An Integrable Deformation of the $AdS_5 \times S^5$ Superstring,''
J.\ Phys.\ A {\bf 47} (2014) 49, 495402
[\arxivlink{1409.1538}].

\bibitem{Sfetsos:2013wia}
K.~Sfetsos,
``Integrable interpolations: From exact CFTs to non-Abelian T-duals,''
Nucl.\ Phys.\ B {\bf 880} (2014) 225
[\arxivlink{1312.4560}].

\bibitem{tse}
A.~A.~Tseytlin,
``On a ``universal'' class of WZW type conformal models,''
Nucl.\ Phys.\ B {\bf 418}, 173 (1994)
[\arxivlink{9311062}].


\bibitem{abf1}
G.~Arutyunov, R.~Borsato and S.~Frolov,
``S-matrix for strings on $\eta$-deformed $AdS_5 \times S^5$,''
JHEP {\bf 1404} (2014) 002
[\arxivlink{1312.3542}].

\bibitem{hrt}
B.~Hoare, R.~Roiban and A.~A.~Tseytlin,
``On deformations of $AdS_n \times S^n$ supercosets,''
JHEP {\bf 1406}, 002 (2014)
[\arxivlink{1403.5517}].


\bi{abf2}
  G.~Arutyunov, R.~Borsato and S.~Frolov,
  ``Puzzles of $\eta$-deformed AdS$_5 \times$ S$^5$,''
  JHEP {\bf 1512}, 049 (2015)
  [\arxivlink{1507.04239}].

\bibitem{Delduc:2013fga} 
  F.~Delduc, M.~Magro and B.~Vicedo,
  ``On classical $q$-deformations of integrable sigma-models,''
  JHEP {\bf 1311}, 192 (2013)
  [\arxivlink{1308.3581}].

\bibitem{ST}
K.~Sfetsos and D.~C.~Thompson,
``Spacetimes for $\lambda$-deformations,''
JHEP {\bf 1412}, 164 (2014)
[\arxivlink{1410.1886}].

\bibitem{ST2}
S.~Demulder, K.~Sfetsos and D.~C.~Thompson,
``Integrable $\lambda$-deformations: Squashing Coset CFTs and $AdS_5 \times S^5$,''
  JHEP {\bf 1507}, 019 (2015)
[\arxivlink{1504.02781}].


\bi{us1}
  B.~Hoare and A.~A.~Tseytlin,
  ``On integrable deformations of superstring sigma models related to $AdS_n \times S^n$ supercosets,''
  Nucl.\ Phys.\ B {\bf 897}, 448 (2015)
  [\arxivlink{1504.07213}].
  
  
  \bibitem{berk} 
N.~Berkovits and J.~Maldacena,
``Fermionic T-Duality, Dual Superconformal Symmetry, and the Amplitude/Wilson Loop Connection,''
JHEP {\bf 0809}, 062 (2008)
[\arxivlink{0807.3196}].
N.~Beisert, R.~Ricci, A.~A.~Tseytlin and M.~Wolf,
``Dual Superconformal Symmetry from $AdS_5 \times S^5$ Superstring Integrability,''
Phys.\ Rev.\ D {\bf 78}, 126004 (2008)
[\arxivlink{0807.3228}].
M. Abbott, J. Murugan, S. Penati, A. Pittelli, D. Sorokin, P. Sundin, J. Tarrant, M. Wolf   and L. Wulff, 
``T-duality of Green-Schwarz superstrings on AdS$_{d} \times$ S$^{d} \times$ M$^{10-2d}$,''
  JHEP {\bf 1512}, 104 (2015)
  [\arxivlink{1509.07678}].

\bi{mts}
R.~R.~Metsaev and A.~A.~Tseytlin,
  ``Type IIB superstring action in AdS(5) x S5 background,''
  Nucl.\ Phys.\ B {\bf 533}, 109 (1998)
  [\arxivlink{hep-th/9805028}].



\bi{bersh}
N.~Berkovits, M.~Bershadsky, T.~Hauer, S.~Zhukov and B.~Zwiebach,
  ``Superstring theory on AdS(2) x S2 as a coset supermanifold,''
  Nucl.\ Phys.\ B {\bf 567}, 61 (2000)
  [\arxivlink{hep-th/9907200}].

  \bi{us2}
  B.~Hoare and A.~A.~Tseytlin,
  ``Type IIB supergravity solution for the T-dual of the $\eta$-deformed AdS$_{5} \times$ S$^{5}$ superstring,''
  JHEP {\bf 1510}, 060 (2015)
  [\arxivlink{1508.01150}].
  
 
\bi{us3}
  G.~Arutyunov, S.~Frolov, B.~Hoare, R.~Roiban and A.~A.~Tseytlin,
  ``Scale invariance of the eta-deformed AdS5 x S5 superstring, T-duality and modified type II equations,''
  Nucl.\ Phys.\ B {\bf 903}, 262 (2016)
  [\arxivlink{1511.05795}].




\bi{holl}
  C.~Appadu and T.~J.~Hollowood,
  ``Beta function of k deformed AdS$_{5}$ x S$^{5}$ string theory,''
  JHEP {\bf 1511}, 095 (2015)
  [\arxivlink{1507.05420}].



\bi{lrt}
O.~Lunin, R.~Roiban and A.~A.~Tseytlin,
``Supergravity backgrounds for deformations of $AdS_n \times S^n$ supercoset string models,''
Nucl.\ Phys.\ B {\bf 891}, 106 (2015)
[\arxivlink{1411.1066}].

\bibitem{Fateev:1992tk}
V.~A.~Fateev, E.~Onofri and A.~B.~Zamolodchikov,
``The Sausage model (integrable deformations of $O(3)$ sigma model),''
Nucl.\ Phys.\ B {\bf 406}, 521 (1993).


\bi{sor}
  D.~Sorokin, A.~Tseytlin, L.~Wulff and K.~Zarembo,
  ``Superstrings in AdS(2)xS(2)xT(6),''
  J.\ Phys.\ A {\bf 44}, 275401 (2011)
  [\arxivlink{1104.1793}].



\bi{hull}
C.~M.~Hull,
``Timelike T duality, de Sitter space, large N gauge theories and topological field theory,''
JHEP {\bf 9807}, 021 (1998)
[\arxivlink{hep-th/9806146}].

\bi{rab}
K.~Bardakci, M.~J.~Crescimanno and E.~Rabinovici,
  ``Parafermions From Coset Models,''
  Nucl.\ Phys.\ B {\bf 344}, 344 (1990).
E.~Witten,
  ``On string theory and black holes,''
  Phys.\ Rev.\ D {\bf 44}, 314 (1991).

\bi{klim}
C.~Klimcik and P.~Severa,
  ``Dressing cosets,''
  Phys.\ Lett.\ B {\bf 381}, 56 (1996)
  [\arxivlink{hep-th/9602162}].
  L.~K.~Balazs, J.~Balog, P.~Forgacs, N.~Mohammedi, L.~Palla and J.~Schnittger,
  ``Quantum equivalence of sigma models related by nonAbelian duality transformations,''
  Phys.\ Rev.\ D {\bf 57}, 3585 (1998)
  [\arxivlink{hep-th/9704137}].

\bibitem{bars} 
  I.~Bars and K.~Sfetsos,
  ``Global analysis of new gravitational singularities in string and particle theories,''
  Phys.\ Rev.\ D {\bf 46}, 4495 (1992)
  [\arxivlink{hep-th/9205037}].
  
 
 
\bibitem{Wulff:2013kga}
  L.~Wulff,
  ``The type II superstring to order $\theta^4$,''
  JHEP {\bf 1307} (2013) 123
  [\arxivlink{1304.6422}].

\bibitem{Cvetic:1999zs} 
  M.~Cvetic, H.~Lu, C.~N.~Pope and K.~S.~Stelle,
  ``T duality in the Green-Schwarz formalism, and the massless / massive IIA duality map,''
  Nucl.\ Phys.\ B {\bf 573}, 149 (2000)
  [\arxivlink{9907202}].

\bibitem{gris} 
  M.~T.~Grisaru, P.~S.~Howe, L.~Mezincescu, B.~Nilsson and P.~K.~Townsend,
  ``N=2 Superstrings in a Supergravity Background,''
  Phys.\ Lett.\ B {\bf 162}, 116 (1985).

 
  
\end{thebibliography}
\end{document}

\bibitem{Fateev:1996ea}
V.~A.~Fateev,
``The sigma model (dual) representation for a two-parameter family of integrable quantum field theories,''
Nucl.\ Phys.\ B {\bf 473}, 509 (1996).
S.~L.~Lukyanov,
``The integrable harmonic map problem versus Ricci flow,''
Nucl.\ Phys.\ B {\bf 865}, 308 (2012)
[\arxivlink{arXiv:1205.3201}].